\documentclass[structabstract]{aa}  
\usepackage{url,cancel,graphics,latexsym,epsfig,graphicx,amssymb,lscape,color,txfonts,natbib}

\renewcommand{\d}[0]{{\rm d}}
\renewcommand{\i}[0]{{\rm i}}
\newcommand{\e}[0]{{\rm e}}
\newcommand{\ave}[1]{\langle #1 \rangle}
\newcommand{\Ave}[1]{\Big\langle #1 \Big\rangle}
\newcommand{\Ref}[1]{(\ref{#1})}
\newcommand{\mat}[1]{{\tens{#1}}}
\newcommand{\pprime}[0]{{\prime\prime}}

\begin{document}

\title{Retrieving the three-dimensional matter power spectrum and
  galaxy biasing parameters from lensing tomography}

\author{Patrick Simon}

\institute{Argelander-Institut f\"ur Astronomie, Universit\"at Bonn, Auf dem
  H\"ugel 71, 53121 Bonn, Germany\\
  \email{psimon@astro.uni-bonn.de}}

\date{Received \today} 

\authorrunning{Patrick Simon}

\titlerunning{Three-dimensional band power spectra from lensing
  tomography}

\def\LaTeX{L\kern-.36em\raise.3ex\hbox{a}\kern-.15em
    T\kern-.1667em\lower.7ex\hbox{E}\kern-.125emX}

\abstract{} 
{With the availability of galaxy distance indicators in weak lensing
  surveys, lensing tomography can be harnessed to constrain the
  three-dimensional (3D) matter power spectrum over a range of
  redshift and physical scale. By combining galaxy-galaxy lensing and
  galaxy clustering, this can be extended to probe the 3D galaxy-matter
  and galaxy-galaxy power spectrum or, alternatively, galaxy biasing
  parameters.}
{To achieve this aim, this paper introduces and discusses minimum
  variance estimators and a more general Bayesian approach to
  statistically invert a set of noisy tomography two-point correlation
  functions, measured within a confined opening angle. Both methods
  are constructed such that they probe deviations of the power
  spectrum from a fiducial power spectrum, thereby enabling both a
  direct comparison of theory and data, and in principle the
  identification of the physical scale and redshift of deviations. By
  devising a new Monte Carlo technique, we quantify the measurement
  noise in the correlators for a fiducial survey, and test the
  performance of the inversion techniques.}
{For a relatively deep \mbox{$200\,\rm deg^2$} survey
  (\mbox{$\bar{z}\sim0.9$}) with 30 sources per square arcmin, the
  matter power spectrum can be probed with \mbox{$3-6\sigma$}
  significance on comoving scales \mbox{$1\lesssim k\,h^{-1}{\rm
      Mpc}\lesssim10$} and \mbox{$z\lesssim0.3$}. For 3 lenses per
  square arcmin, a significant detection ($\sim10\sigma$) of the
  galaxy-matter power spectrum and galaxy power spectrum is attainable
  to relatively high redshifts ($z\lesssim0.8$) and over a wider
  $k$-range.  Within the Bayesian framework, all three power spectra
  are easily combined to provide constraints on 3D galaxy biasing
  parameters. Therein, weak priors on the galaxy bias improve
  constraints on the matter power spectrum.}
{A shear tomography analysis of weak-lensing surveys in the near
  future promises fruitful insights into both the effect of baryons on
  the nonlinear matter power spectrum at \mbox{$z\lesssim0.3$} and
  galaxy biasing (\mbox{$z\lesssim0.5$}). However, a proper treatment
  of the anticipated systematics, which are not included in the mock
  analysis but discussed here, is likely to reduce the signal-to-noise
  ratio in the analysis such that a robust assessment of the 3D matter
  power spectrum probably requires a survey area of at least
  $\sim10^3\,\rm deg^2$. To investigate the matter power spectrum at
  redshift higher than $\sim0.3$, an increase in survey area is
  mandatory.}


\maketitle


\section{Introduction}

In the framework of the standard model of cosmology
\citep[e.g.,][]{2003moco.book.....D}, structure in the matter density
field evolves over time under the influence of gravitational collapse
from a relatively homogeneous state, as for instance observed in the
small temperature fluctuations of the cosmic microwave background
\citep[CMB;][]{2011ApJS..192...18K}, to the highly structured matter
density field of today. The main driver of the structure formation is
the postulated dark matter, which is the dominant matter
component. According to the model, dark matter only interacts by means
of either the weak nuclear force or gravity. Moreover, the first
galaxies were formed from the subdominant primordial baryonic gas
component, building stars within the gravitational potential wells of
the dark matter \citep[e.g.,][]{2010gfe..book.....M}, and evolved
later on.

To quantify the state of the structure formation process at a given
cosmic time or a radial comoving distance $\chi$, the amplitude of the
fluctuations
\begin{equation}
  \delta_{\rm m}(\vec{x},\chi)=
  \frac{\rho_{\rm m}(\vec{x},\chi)-\bar{\rho}_{\rm m}(\chi)}
  {\bar{\rho}_{\rm m}(\chi)}
\end{equation} 
in the matter density field $\rho_{\rm m}(\vec{x},\chi)$ relative to
the mean matter density $\bar{\rho}_{\rm m}(\chi)$, at a comoving
position $\vec{x}$, is expressed by the matter power spectrum $P_{\rm
  m}(k,\chi)$ \citep[e.g.,][]{2002sgd..book.....M}
\begin{equation}
  \Ave{\tilde{\delta}_{\rm m}(\vec{k},\chi)\tilde{\delta}_{\rm m}(\vec{k}^\prime,\chi)}=
  (2\pi)^3\delta_{\rm D}^{(3)}(\vec{k}+\vec{k}^\prime)P_{\rm m}(k,\chi)\;,
\end{equation}
where 
\begin{equation}
  \tilde{\delta}_{\rm m}(\vec{k},\chi)=
  \int\d^3x\,\delta_{\rm m}(\vec{x},\chi)\e^{-\i\vec{x}\cdot\vec{k}}
\end{equation}
are the amplitudes of the density modes with wave-number $|\vec{k}|$
or wave length $2\pi/|\vec{k}|$, $\delta^{(3)}_{\rm D}(\vec{x})$ is
the Dirac delta function, and $\ave{\ldots}$ denotes the ensemble
average over all realisations of the matter density field. In the
cosmological context, density fields are statistically homogeneous and
isotropic.

The matter power spectrum contains a wealth of information directly
related to the physics of the matter components in the Universe as
well as the primordial density fluctuations \citep[][and references
therein]{1999coph.book.....P} and information about possible
modifications to Einstein's general relativity that may affect the
growth of the density fluctuations
\citep{2001PhRvD..64h3004U,2011arXiv1106.2476C}.  While the matter
power spectrum at redshift \mbox{$z\sim1000$} is relatively
well-understood theoretically and known empirically from CMB studies
\citep{2011ApJS..192...18K}, the power spectrum in the more recent
Universe on quasi- to nonlinear scales \mbox{$k\gtrsim1\,h\rm
  Mpc^{-1}$} is less well-known: On the theoretical side, the
magnitude of the influence of galaxies and baryons seems to be unclear
but probably significant \citep{2004ApJ...616L..75Z,
  2006ApJ...640L.119J, 2010PhRvD..81b3524S,
  2011arXiv1105.1075S,2011MNRAS.415.3649V}, although predictions for
dark-matter-only scenarios are accurate
\citep{1996MNRAS.280L..19P,Smith03,2010ApJ...715..104H}. On the
empirical side, detailed studies of the power spectrum in this regime
rely mostly on tracers \citep{2004ApJ...606..702T}, such as galaxies
or absorbers in the Lyman-$\alpha$ forest, which have an uncertain
relation to the underlying matter density field. This provides a
strong motivation to directly measure either the spatial matter power
spectrum or the spatial matter correlation function.

One purpose of this paper is to propose estimators of the matter power
spectrum based on the weak gravitational lensing effect to allow a
direct measurement of $P_{\rm m}(k,\chi)$, especially in the nonlinear
regime. Weak gravitational lensing \citep[][for a
review]{2006glsw.conf..269S} is a tool to study the large-scale
distribution of matter by its effect on the shape of distant galaxy
images (``sources''). Images of galaxies are weakly distorted
(sheared) by the bending of light passing by intervening matter
density fluctuations (``lenses''). As the effect is only sensitive to
the fluctuations in the gravitational field of intervening matter, it
can be employed to probe the matter density field without further
knowledge of its physical properties and without the usage of
tracers. This makes it an excellent probe to explore the elusive dark
matter component or the general composition of matter on cosmological
scales (see for example \citealt{Clowe06_bullet} or
\citealt{2009A&A...500..657T}).

As weak lensing is sensitive to all intervening matter, it naturally
measures the projected matter density fluctuations within radial
cylinders. Known lensing-based estimators of the matter power spectrum
are therefore essentially two-dimensional by either estimating the
projected matter fluctuations on the sky
\citep{1991MNRAS.251..600B,1991ApJ...380....1M,1992ApJ...388..272K,1998MNRAS.296..873S,2001ApJ...554...67H,2002A&A...396....1S,2002ApJ...567...31P}
or translating this into a spatial power spectrum at one effective
distance \citep{2003MNRAS.346..994P}. With the availability of
distance indicators for the sources, one has started to exploit the
change in the weak lensing shear $\gamma(\chi_{\rm s})$ as a function
of comoving source distance $\chi_{\rm s}$. For a fixed lens distance
$\chi_{\rm d}$, this is expressed by
\begin{equation}
  \gamma(\chi_{\rm s})\propto
\frac{f_{\rm k}(\chi_{\rm s}-\chi_{\rm d})}{f_{\rm   k}(\chi_{\rm s})}
\end{equation}
with $f_{\rm k}(\chi)$ being the comoving angular diameter
distance. The weight of a lens in the total shear signal, including
lenses at all distances, therefore varies with a varying source
distance. In a cosmic shear tomography analysis, this is utilised to
provide improved constraints on cosmological parameters
\citep{2002PhRvD..66h3515H,2007ApJS..172..239M,schrabback2010,2011MNRAS.413.2923K}
or to map the three-dimensional (3D) matter density field
\citep{2002PhRvD..66f3506H,MasseyNat,2011ApJ...727..118V,2011arXiv1109.0932S,2011arXiv1111.6478L}
by analysing the lensing signal for a series of source distance
slices.

Weak lensing also offers the opportunity to investigate the
distribution of cosmological objects in relation to the matter density
field. As in the case of the matter power spectrum, one defines the
clustering power spectrum of galaxies
\begin{equation}
  \label{eq:galpower}
  \Ave{\tilde{\delta}_{\rm g}(\vec{k},\chi)\tilde{\delta}_{\rm g}(\vec{k}^\prime,\chi)}=
  (2\pi)^3\delta_{\rm D}^{(3)}(\vec{k}+\vec{k}^\prime)
  \left(P_{\rm
    g}(k,\chi)+\frac{1}{\bar{n}_{\rm g}}\right)
\end{equation}
and the cross-correlation power of the matter and galaxy distribution
\begin{equation}
  \Ave{\tilde{\delta}_{\rm g}(\vec{k},\chi)\tilde{\delta}_{\rm m}(\vec{k}^\prime,\chi)}=
  (2\pi)^3\delta_{\rm D}^{(3)}(\vec{k}+\vec{k}^\prime)P_{\delta\rm g}(k,\chi)\;,
\end{equation}
based on the galaxy number density contrast $\delta_{\rm g}=n_{\rm
  g}/\bar{n}_{\rm g}-1$, where $n_{\rm g}$ is the number density of
galaxies, and $\bar{n}_{\rm g}$ their mean number density. We note
that the definition of $P_{\rm g}(k,\chi)$ separates the Poisson shot
noise contribution, originating from the discreteness of the galaxies,
from the number density fluctuations
\citep[e.g.,][]{1999coph.book.....P}. For galaxies faithfully tracing
the distribution of matter, all three power spectra
$P_\delta(k,\chi)$, $P_{\delta\rm g}(k,\chi)$, and $P_{\rm g}(k,\chi)$
are identical. More generally, however, we have to assume that
galaxies are imperfect tracers, i.e., biased tracers, of the matter
distribution. Galaxy biasing is commonly expressed at the power
spectrum level by the biasing functions $b(k,\chi)$ and $r(k,\chi)$
\citep[e.g.,][]{1998ApJ...500L..79T}:
\begin{eqnarray}
  P_{\delta\rm g}(k,\chi)&=&b(k,\chi)r(k,\chi)P_\delta(k,\chi)\;,\\
  P_{\rm g}(k,\chi)&=&b^2(k,\chi)P_\delta(k,\chi)\;.
\end{eqnarray}
Measuring galaxy biasing addresses the question of galaxy formation
and evolution as deviations of galaxy clustering from (dark) matter
clustering -- as reflected by the biasing functions with
\mbox{$b,r\ne1$} -- constrain galaxy models
\citep{2001ApJ...558..520Y,2004ApJ...601....1W,2005Natur.435..629S}.

Non-parametric weak gravitational-lensing methods have been proposed
\citep{1998A&A...334....1V,1998ApJ...498...43S,2003MNRAS.346..994P}
and applied
\citep{2002ApJ...577..604H,2003MNRAS.346..994P,2007A&A...461..861S} to
assess the galaxy biasing for an effective radial distance, but to
date without exploiting tomography. The formalism in this paper is
kept general enough to constrain the galaxy biasing functions
$b(k,\chi)$ and $r(k,\chi)$ or, alternatively, the power spectra
$P_{\delta\rm g}(k,\chi)$ and $P_{\rm g}(k,\chi)$ as well.

To enable the employment of this new tomography technique, the
appropriate lensing data will soon be available.  Recent, ongoing, and
prospective lensing surveys -- such as the Canada-France-Hawaii Legacy
Survey
(\mbox{CFHTLS}\footnote{\url{http://www.cfht.hawaii.edu/Science/CFHLS/}}),
the Panoramic Survey Telescope and Rapid Response System surveys
(\mbox{Pan-STARRS}\footnote{\url{http://pan-starrs.ifa.hawaii.edu}}),
the Dark Energy Survey
(\mbox{DES}\footnote{\url{http://www.darkenergysurvey.org}}), and the
Kilo-Degree Survey
(\mbox{KiDS}\footnote{\url{http://www.astro-wise.org/projects/KIDS/}})
-- will provide between hundreds and thousands of square degrees of
shear data endowed with galaxy redshifts. With data sets of this
magnitude in hand, we can expect to extract the 3D matter power
spectrum and biasing parameters of a selected galaxy population, as
discussed in the following.

The paper is structured as follows. Sect. \ref{sect:quadratic} derives
a set of minimum variance estimators, which, when applied to
measurements of angular correlation functions detailed below, yield
constraints on the 3D power spectra $P_\delta(k,\chi)$, $P_{\delta\rm
  g}(k,\chi)$, and $P_{\rm g}(k,\chi)$. In
Sect. \ref{sect:mvforecast}, the estimators are applied to a fiducial
survey (Sect. \ref{sect:survey}) to forecast their performance. The
estimators are improved within a Bayesian analysis in
Sect. \ref{sect:bayes}.  Sect. \ref{sect:bayesforecast} applies the
Bayesian methodology to the fiducial survey. Finally, in
Sect. \ref{sect:discuss} we discuss the results and draw general
conclusions.

As a fiducial cosmology, this paper uses a $\Lambda$CDM model
(adiabatic fluctuations) with matter density parameter
\mbox{$\Omega_{\rm m}=0.27$}, of which baryons are given by
\mbox{$\Omega_{\rm b}=0.046$}, a cosmological constant
\mbox{$\Omega_{\Lambda}=1-\Omega_{\rm m}$}, and a Hubble parameter
\mbox{$H_{0}=h\,100\,{\rm km}\,{\rm s}^{-1}\,{\rm Mpc}^{-1}$}. Length
scales are quoted for \mbox{$h=1$}, while the shape parameter of the
dark matter clustering adopts \mbox{$h=0.704$}. The normalisation of
the matter fluctuations within a sphere of radius $8\,h^{-1}\rm Mpc$
at redshift zero is assumed to be $\sigma_8=0.81$. For the spectral
index of the primordial matter power spectrum, we use $n_{\rm
  s}=0.96$.

\section{Minimum variance estimators of band power spectra}
\label{sect:quadratic}

For extensive recent reviews of weak gravitational lensing and the
weak lensing formalism, we refer the reader to
\citet{2003astro.ph..5089V} or \citet{2006glsw.conf..269S}.

As we show in the following, fixing the angular diameter distance
$f_{\rm k}(\chi)$, the matter density parameter $\Omega_{\rm m}$, and
the relation between comoving radial distance and redshift
\begin{equation}
  \chi(z)=\int_0^z\frac{c\d z^\prime}{H(z^\prime)}
\end{equation}
within a reference fiducial cosmology (where $H(z)$ is the Hubble
parameter at redshift $z$), allows us to find a simple linear relation
between the 3D power spectra and the angular correlation functions as
long as lens-lens couplings are negligible. This is assumed for the
scope of this paper.

The lens and source catalogues from a galaxy lensing survey split into
$N_{\rm source}$ and $N_{\rm lens}$ subsamples, each covering
different radial distance regimes. Quantities computed from the $i$th
subsample are referred to in the following by a superindex as in
``$q^{(i)}$''. Likewise, quantities computed from a pair of subsamples
$i$ and $j$ are denoted by ``$q^{(ij)}$''. By tomography, we mean a
set of angular two-point correlation functions obtained from pairs of
galaxy subsamples. The pair combinations include galaxies from both
different and the same redshift bins.

\subsection{Matter power spectrum}
\label{sect:pm}

Here we start with shear-shear correlations between sources
\citep[e.g.,][and references therein]{2006glsw.conf..269S}
\begin{equation}
  \xi_\pm^{(ij)}(|\vec{\Delta\theta}|)=
  \Ave{\gamma^{(i)}_{\rm t}(\vec{\theta})\gamma^{(j)}_{\rm
      t}(\vec{\theta}^\prime)}\pm
  \Ave{\gamma^{(i)}_\times(\vec{\theta})\gamma^{(j)}_\times(\vec{\theta}^\prime)}\;,
\end{equation} 
where $\gamma^{(i)}_\times$ and $\gamma^{(i)}_{\rm t}$ denote the
cross and tangential shear component of sources in the $i$th
subsample, which are both evaluated relative to the line
$\vec{\Delta\theta}=\vec{\theta}^\prime-\vec{\theta}$ connecting the
two sources in directions $\vec{\theta}$ and
$\vec{\theta}^\prime$. For shear fields and positions on the (flat)
sky, we employ the complex notation, which is widely used in the
literature. The source catalogue is split into $N_{\rm source}$
subsamples with known radial distributions $p_\chi^{(i)}(\chi)$ that
define the probability of finding a source within the comoving
distance interval $[\chi,\chi+\d\chi]$ by $p_\chi^{(i)}(\chi)\d\chi$.

Using Limber's approximation \citep{1992ApJ...388..272K}, the
$\xi_\pm^{(ij)}(\theta)$ in the shear tomography can be related to the
3D matter power spectrum as in \citep[e.g.,][]{2004A&A...417..873S}
\begin{eqnarray}
  \label{eq:xipm}
  \lefteqn{\xi_\pm^{(ij)}(\theta)=}\\
  &&\nonumber
  \frac{9H_0^4\Omega_{\rm m}^2}{4c^4}
  \int_0^{\chi_{\rm h}}
  \!\!\!\!\int_0^\infty\frac{\d\chi\d\ell\,\ell}{2\pi}
  \frac{\overline{W}^{(i)}(\chi)\overline{W}^{(j)}(\chi)}{a(\chi)}
  J_{0,4}(\ell\theta)P_\delta\left(\frac{\ell}{f_{\rm k}(\chi)},\chi\right)\;,
\end{eqnarray}
where the average lensing efficiency is defined as
\begin{equation}
  \overline{W}^{(i)}(\chi)=
  \int_\chi^{\chi_{\rm h}}\d\chi^\prime
  p_\chi^{(i)}(\chi^\prime)
  \frac{f_{\rm k}(\chi^\prime-\chi)}{f_{\rm k}(\chi^\prime)}\;.
\end{equation}
We have also introduced $a(\chi)$ as the cosmic scale factor at
comoving distance $\chi$, the $n$th-order Bessel function of the first
kind $J_n(x)$, and the comoving radius of the observable Universe
$\chi_{\rm h}$.

Moving on to expand the 3D matter power spectrum $P_\delta(k,\chi)$ as
a linear combination of basis functions, we define a 2D grid with grid
points $(k_m,\chi_n)$ and $m\in\{1,\ldots,N_k+1\},
n\in\{1,\ldots,N_z+1\}$. The grid positions are sorted,
\mbox{$k_m<k_{m+1}$} and \mbox{$\chi_n<\chi_{n+1}$}. The grid points
mark the interval limits of a band power spectrum. To focus on
deviations from a model power spectrum, the band powers are defined
relative to a fiducial matter power spectrum $P^{\rm
  fid}_\delta(k,\chi)$ that describes the expected variation in the
power within the bands or regions outside the grid
\begin{eqnarray}
  \label{eq:Pmansatz}
  P_\delta(k,\chi)&=&
  P^{\rm fid}_\delta(k,\chi)\left(
    1+\sum_{n=1}^{N_z}\sum_{m=1}^{N_k}
    H_{mn}(k,\chi)\left[f_{\delta,mn}-1\right]\right)\;,
\end{eqnarray}
where $f_{\delta,mn}$ are factors allowing a deviation from the
fiducial band powers, and
\begin{equation}
  H_{mn}(k,\chi):=\left\{
    \begin{array}{ll}
      1 & {\rm if}~k\in[k_m,k_{m+1}[~{\rm
        and}~\chi\in[\chi_n,\chi_{n+1}[\\
      0 & {\rm otherwise}
    \end{array}\right.
\end{equation}
defines a 2D top hat function. Therefore, $P_\delta(k,\chi)$ is
approximated by in total $N_kN_z$ free parameters $f_{\delta,\rm
  mn}$. The boundaries $(k_1,k_{N_k+1})$ and $(\chi_1,\chi_{N_z+1})$
confine the range that is allowed to be modified in comparison to the
fiducial model. According to Eq. \Ref{eq:Pmansatz}, outside this
region one has only \mbox{$P_\delta(k,\chi)=P^{\rm
    fid}_\delta(k,\chi)$}. Inside the region, a value of
\mbox{$f_{\delta,mn}=1$} uses either the power expected from the
fiducial model or a different power amplitude. In this paper, we focus
in particular on evaluating the coefficients $f_{\delta, mn}$ and
similar coefficients for other power spectra we introduce below. We
note that by setting \mbox{$P^{\rm fid}_\delta(k,\chi)\equiv1$} and
the grid being sufficiently large, we approximate the matter power
spectrum by constant powers within the band with absolute amplitudes
$f_{\delta,mn}$. Adopting a concrete fiducial model spectrum as a
reference may, however, be more advantageous as it implements trends
in $k$ and $\chi$ within the (broad) bands as expected from theory. If
the trends (not the absolute amplitudes) are more or less realistic,
we can expect the factors $f_{\delta,mn}$ to be only slowly changing
functions of either $k$ or $\chi$. Moreover, the fiducial model makes
the reconstruction more stable as it specifies the power spectrum
within a $k$-regime that is only poorly constrained owing to the
limited angular range that is covered by the angular correlation
function. For the scope of this paper, where \mbox{$f_{\delta,\rm
    mn}=1$} is constant throughout, we keep the number of bins with
\mbox{$N_z=5$} (linear) and \mbox{$N_k=10$} (log-bins) relatively
small.

By virtue of this band power approximation, Eq.  \Ref{eq:xipm} can be
cast into the linear form
\begin{equation}
  \label{eq:xipmband} 
  \xi_\pm^{(ij)}(\theta)=
  \sum_{n=1}^{N_z}\sum_{m=1}^{N_k}
  X^{(ij)}_\pm(\theta;m,n)f_{\delta,mn}+
  \xi_{\pm,\rm fid}^{(ij)}(\theta)\;,
\end{equation}
where we utilise the basis functions
\begin{eqnarray}
  \label{eq:xipmbasis}
  \lefteqn{X^{(ij)}_\pm(\theta;m,n):=
    \frac{9H_0^4\Omega_{\rm m}^2}{8\pi c^4\theta^2}\,\times}\\
  &&\nonumber
  \int_{\chi_n}^{\chi_{n+1}}\d\chi
  \frac{\overline{W}^{(i)}(\chi)\overline{W}^{(j)}(\chi)}{a(\chi)}
  \int\limits_{k_m f_{\rm k}(\chi)\theta}^{k_{m+1}f_{\rm k}(\chi)\theta}\d s\,s\,
  J_{0,4}(s)\,P^{\rm fid}_\delta\left(\frac{s}{f_{\rm k}(\chi)\theta},\chi\right)\;;
\end{eqnarray}
the Bessel function $J_0$ is used for $X^{(ij)}_+$, whereas $J_4$ has
to be used in the case of $X^{(ij)}_-$. A short description of the
numerical evaluation of this integral and similar integrals to follow
can be found in Appendix \ref{sect:calculus}. The fiducial correlation
function $\xi_{\pm,\rm fid}^{(ij)}$ accounts for any possible
invariable contributions from the $(k,\chi)$-plane that are not
covered by the grid
\begin{eqnarray}
  \lefteqn{\xi_{\pm,\rm fid}^{(ij)}(\theta):=
    \frac{9H_0^4\Omega_{\rm m}^2}{8\pi c^4\theta^2}\,\times}\\
  &&\nonumber
  \int_{0}^{\chi_{\rm h}}\d\chi
  \frac{\overline{W}^{(i)}(\chi)\overline{W}^{(j)}(\chi)}{a(\chi)}
  \int\limits_{0}^{\infty}\d s\,s\,
  J_{0,4}(s)\,
  \cancel{P}^{\rm fid}_\delta\left(\frac{s}{f_{\rm k}(\chi)\theta},\chi\right)\;,
\end{eqnarray}
where
\begin{equation}
\cancel{P}^{\rm fid}_\delta\left(k,\chi\right):=
\left\{
\begin{array}{ll}
  P^{\rm fid}_\delta(k,\chi) & ,~{\rm if}~\sum\limits_{m,n=1}^{N_z,N_k}H_{mn}(k,\chi)=0\\
  0 & ,~{\rm otherwise}
\end{array}
\right.\;.
\end{equation}
For sufficiently large grids as in the mock analysis below, however,
we find this contribution to be negligible.

In practise, one obtains a measurement of $\xi_\pm^{(ij)}(\theta_l)$
for a series of $l\in\{1,\ldots,N_\theta\}$ $\theta$-bins and
\mbox{$N_{\rm source}(N_{\rm source}+1)/2$} combinations of source
subsamples $(ij)$. After arranging the series of data points as one
compact shear tomography data vector $\vec{\xi}$ with $N_\theta N_{\rm
  source}(N_{\rm source}+1)$ elements, the set of equations in
Eq. \Ref{eq:xipmband} for all index pairs $(ij)$ can be written as
\begin{equation}
  \label{eq:xiXpm}
  \vec{\xi}=\mat{X}\vec{f}_\delta+\vec{\xi}_{\rm fid}\;.
\end{equation}
Here, the band power factors $f_{\delta.mn}$, compiled within the
vector $\vec{f}_\delta$, and the matrix elements of $\mat{X}$,
consisting of the $X_\pm^{(ij)}(\theta_l;m,n)$, are arranged to comply
with the structure of $\vec{\xi}$. The constant vector $\vec{\xi}_{\rm
  fid}$ consists of the fiducial correlation-function values
$\xi^{(ij)}_{\pm,\rm fid}(\theta_l)$.

For a given noisy $\vec{\xi}$ from observation, this equation has to
be inverted with respect to $\vec{f}_\delta$. As an estimator for
$\vec{f}_\delta$, we suggest a minimum variance estimator
\citep[cf.][]{1995ApJ...449..446Z}
\begin{equation}
  \label{eq:xipmest}
  \hat{\vec{f}}_\delta=
  [\mat{X}^{\rm t}\mat{N}_\xi^{-1}\mat{X}]^{-1}\mat{X}^{\rm
    t}\mat{N}^{-1}_\xi
  \left(\vec{\xi}-\vec{\xi}_{\rm fid}\right)\;,
\end{equation}
which minimises the residual
$\ave{||\hat{\vec{f}}_\delta-\vec{f}_\delta)||^2}$, the average over
all noise realisations, where $||\ldots||$ is the Euclidean norm.
Compared to the simple inverse $\mat{X}^{-1}$, the estimator can also
cope if the system of linear equations in Eq. \Ref{eq:xiXpm} is
over-determined. In this case, redundant information is optimally
combined based on the noise covariance
\begin{equation}
  \mat{N}_\xi:=
  \Ave{\left(\vec{\xi}-\ave{\vec{\xi}}\right)
    \left(\vec{\xi}-\ave{\vec{\xi}}\right)^{\rm t}}\;.
\end{equation}
The covariance of the estimator is
\begin{equation}
  \label{eq:noisedelta}
  \mat{N}_\delta=
  \Ave{\vec{\hat{f}}_\delta\vec{\hat{f}}^{\rm t}_\delta}-
    \Ave{\vec{\hat{f}}_\delta}\Ave{\vec{\hat{f}}^{\rm t}_\delta}=
  [\mat{X}^{\rm t}\mat{N}_\xi^{-1}\mat{X}]^{-1}\;,
\end{equation} 
which can be used to quantify the error and the correlation of errors
in the estimate. Furthermore, the estimator is unbiased when
$\vec{\xi}_{\rm fid}$ is correct.

The estimator will fail when the matrix product in the square brackets
of $\mat{N}_\delta$ is singular, which happens when the problem is
ill-conditioned. In this context, ill-conditioned can mean that either
$\vec{f}_\delta$ comprises a redshift or length-scale range that is
unconstrained by $\vec{\xi}$, or the impact of different bands on
$\vec{\xi}$ is entirely degenerate. One usually has to modify the
$(k_m,\chi_n)$ band boundaries, the redshift slicing, or the fiducial
power spectrum $P^{\rm fid}_\delta(k,\chi)$ accordingly to remove the
singularity.  To evade the problem of the undefined estimator, we use
a singular-value decomposition \citep[SVD;][]{1992nrca.book.....P} for
the inversion of $\mat{X}^{\rm t}\mat{N}_\xi^{-1}\mat{X}$ in order to
find a pseudo-inverse. The pseudo-inverse $\mat{A}^+$ of a matrix
$\mat{A}$ minimises the matrix norm $||\mat{A}\mat{A}^+-\mat{1}||$,
where $\mat{1}$ is the unity matrix. For a regular matrix $\mat{A}$,
the straightforward solution is $\mat{A}^+=\mat{A}^{-1}$. For a
singular $\mat{A}$, on the other hand, a solution
$\vec{x}^+=\mat{A}^+\vec{b}$ is constrained to be as close as possible
to $\mat{A}\vec{x}=\vec{b}$.

For the problem at hand, the pseudo-inverse can yield a biased
estimator for the unconstrained or degenerate parts of
$\vec{f}_\delta$. To determine the biased elements in
$\hat{\vec{f}}_\delta$, we use as reference
\begin{equation}
  \label{eq:biasest}
  \vec{f}^{\rm ref}_\delta=
  [\mat{N}_\delta^{-1}]^+\mat{N}_\delta^{-1}\vec{1}\;,
\end{equation}
where $\vec{1}$ is a vector with the same number of elements as
$\vec{f}_\delta$ but with all elements set to unity; this is the
estimator in Eq. \Ref{eq:xipmest} applied to the data vector
$\vec{\xi}_{\rm test}=\mat{X}\vec{1}+\vec{\xi}_{\rm fid}$ that is
expected from the fiducial reference model.  The elements in
$\vec{f}^{\rm ref}_\delta$ that are unity are either unbiased or
biased.

\subsection{Galaxy-matter cross-correlation power spectrum}

A line of reasoning similar to that in Sect. \ref{sect:pm} also
applies to the galaxy-matter cross-correlation power spectrum
$P_{\delta\rm g}(k,\chi)$. A projection of the 3D matter-galaxy power
spectrum $P_{\delta\rm g}$ is measured by estimating the mean
tangential shear $\overline{\gamma}_{\rm t}$ about a sample of lenses
with a number density contrast
\begin{equation}
  \kappa_{\rm g}(\vec{\theta})=
  \frac{n_{\rm g}(\vec{\theta})}{\overline{N}_{\rm g}}-1
\end{equation}
 on the sky, namely \citep{2001PhR...340..291B}
\begin{equation}
  \bar{\gamma}_{\rm t}(|\vec{\Delta\theta}|)+
  {\rm i}\bar{\gamma}_\times(|\vec{\Delta\theta}|)=
  -\e^{-2\i\phi}\Ave{\kappa_{\rm g}(\vec{\theta}^\prime)\gamma(\vec{\theta})}\;,
\end{equation}
where $\overline{\gamma}_{\rm t}(\Delta\theta)$ (mean tangential
shear) denotes the mean shear around a lens at separation
$\Delta\theta$ in a reference frame rotated by the angle defined by
the polar angle $\phi$ of
$\vec{\Delta\theta}:=\vec{\theta}-\vec{\theta}^\prime$,
$\bar{\gamma}_\times$ is the mean cross-shear component (mean cross
shear), which cannot be generated to lowest order by gravitational
lensing and may be used as reliable indicator of systematics, and
$n_{\rm g}(\vec{\theta})$ denotes the lens number density on the sky
in the direction of $\vec{\theta}$ and $\overline{N}_{\rm
  g}=\ave{n_{\rm g}(\vec{\theta})}$ the mean lens number density. The
mean tangential or cross shear do not rely on $\phi$ owing to the
statistical isotropy of the fields.

For galaxy samples sliced in redshift with sources from the $i$th
sample and lenses from the $j$th sample, the relation of
$\bar{\gamma}^{(ij)}_{\rm t}(\theta)$ to the 3D matter-galaxy
cross-power spectrum is \citep[e.g.,][]{2007A&A...461..861S}
\begin{eqnarray}
  \label{eq:gammat}
  \lefteqn{\overline{\gamma}_{\rm t}^{(ij)}(\theta)=}\\
  &&\nonumber
  \frac{3H_0^2\Omega_{\rm m}}{2c^2}
  \int_0^{\chi_{\rm
      h}}
  \!\!\!\!\int_0^\infty
  \frac{\d\chi\d\ell\,\ell}{2\pi}
  \frac{\overline{W}^{(i)}(\chi)q_\chi^{(j)}(\chi)}{f_{\rm
      k}(\chi)a(\chi)}
  J_2(\ell\theta)
  P_{\delta\rm g}\left(\frac{\ell}{f_{\rm k}(\chi)},\chi\right)\;. 
\end{eqnarray}
As before with the sources, we need to know the radial (probability)
distribution $q^{(j)}_\chi(\chi)$ for each lens slice.

If we fix the fiducial cosmological model and adopt for $P_{\delta\rm
  g}(k,\chi)$ the previous band power Ansatz in Eq. \Ref{eq:Pmansatz}
with free scaling parameters $f_{{\delta\rm g},mn}$ relative to a
fiducial power spectrum, we now obtain
\begin{equation}
  \overline{\gamma}_{\rm t}^{(ij)}(\theta)=
  \sum_{n=1}^{N_z}\sum_{m=1}^{N_k}
  Y^{(ij)}(\theta;m,n)f_{{\delta\rm g},mn}+
  \overline{\gamma}_{\rm t,fid}^{(ij)}(\theta)
\end{equation}
with the basis functions
\begin{eqnarray}
  \label{eq:gammatbasis}
  \lefteqn{Y^{(ij)}(\theta;m,n)= 
    \frac{3H_0^2\Omega_{\rm m}}{4\pi c^2\theta^2}\,\times}\\
  &&\nonumber
  \int_{\chi_n}^{\chi_{n+1}}\d\chi
  \frac{\overline{W}^{(i)}(\chi)q_\chi^{(j)}(\chi)}{f_{\rm
      k}(\chi)a(\chi)}
  \int\limits_{k_mf_{\rm k}(\chi)\theta}^{k_{m+1}f_{\rm k}(\chi)\theta}\d s\,s\,
  J_2(s)\,
  P^{\rm fid}_\delta\left(\frac{s}{f_{\rm k}(\chi)\theta},\chi\right)\;,
\end{eqnarray}
and the invariable terms
\begin{eqnarray}
  \lefteqn{\overline{\gamma}_{\rm t,fid}^{(ij)}(\theta):=
    \frac{3H_0^2\Omega_{\rm m}}{4\pi c^2\theta^2}\,\times}\\
  &&\nonumber
  \int_{0}^{\chi_{\rm h}}\d\chi
  \frac{\overline{W}^{(i)}(\chi)q_\chi^{(j)}(\chi)}{f_{\rm
      k}(\chi)a(\chi)}
  \int\limits_{0}^{\infty}\d s\,s\,
  J_2(s)\,
  \cancel{P}^{\rm fid}_\delta\left(\frac{s}{f_{\rm k}(\chi)\theta},\chi\right)\;.
\end{eqnarray}
This system of equations is cast into a more compact form
\begin{equation}
  \overline{\vec{\gamma}}_{\rm t}=
  \mat{Y}\vec{f}_{\delta\rm g}+\overline{\vec{\gamma}}_{\rm t,fid}
\end{equation}
for a series of measurements $\overline{\gamma}_{\rm
  t}^{(ij)}(\theta_l)$ and the band power-spectrum coefficients
$f_{{\delta\rm g},mn}$ appropriately arranged inside the vectors
$\overline{\vec{\gamma}}_{\rm t}$ ($N_\theta N_{\rm lens} N_{\rm
  source}$ elements) and $\vec{f}_{\delta\rm g}$ ($N_kN_z$ elements),
respectively. As in the foregoing section, we place all values
$\overline{\gamma}_{\rm t,fid}^{(ij)}(\theta_l)$ into a constant
offset vector, which is here $\overline{\vec{\gamma}}_{\rm t,fid}$.

Following from this linear relation, a minimum variance estimator for
$\vec{f}_{\delta\rm g}$ is now given by
\begin{equation}
  \label{eq:gammatest}
  \vec{\hat{f}}_{\delta\rm g}=
  \big[\mat{N}_{\delta\rm g}^{-1}\big]^+\mat{Y}^{\rm
    t}\mat{N}_{\gamma}^{-1}
  \left(\overline{\vec{\gamma}}_{\rm t}-\overline{\vec{\gamma}}_{\rm t,fid}\right)\;,
\end{equation}
where
\begin{equation}
\label{eq:noisedeltag}
\mat{N}_{\delta\rm g}^{-1}=
\mat{Y}^{\rm t}\mat{N}_\gamma^{-1}\mat{Y}
\end{equation}
is the inverse covariance of the estimator and $\mat{N}_\gamma$ is the
noise covariance of the galaxy-galaxy lensing measurement. We note
that we have chosen the fiducial power spectrum to be identical to the
fiducial 3D matter power spectrum in the last section for
convenience. As lenses are probably biased with respect to the matter
distribution, we should expect $f_{{\delta\rm g},mn}$ in general to
differ from unity. By construction, $f_{\delta{\rm g},mn}$ is an
average of $r(k,\chi)b(k,\chi)$ for the lenses over the corresponding
band relative to the \emph{fiducial} matter power spectrum. By taking
ratios with respect to $f_{\delta,mn}$, we can, however, estimate
$r(k,\chi)b(k,\chi)$ with respect to the true matter power, as done in
the Bayesian analysis below.

\subsection{Galaxy power spectrum}

Finally, we describe our list of minimum variance estimators by
considering the angular clustering of the lenses \citep{peebles80}
\begin{equation}
  \omega^{(ij)}(|\vec{\Delta\theta}|)=
  \Ave{\kappa^{(i)}_{\rm g}(\vec{\theta}^\prime)
    \kappa^{(j)}_{\rm g}(\vec{\theta})}\;,
\end{equation}
which probes the spatial clustering of lenses from the $i$th relative
to the $j$th sample \citep[e.g.,][]{2007A&A...461..861S}
\begin{equation}
  \label{eq:omega}
  \omega^{(ij)}(\theta)=
  \int_0^{\chi_{\rm h}}
  \!\!\!\!\int_0^\infty
  \frac{\d\chi\d\ell\,\ell}{2\pi}
  \frac{q^{(i)}_\chi(\chi)q^{(j)}_\chi(\chi)}{f^2_{\rm k}(\chi)}
  J_0(\ell\theta)P_{\rm g}\left(\frac{\ell}{f_{\rm k}(\chi)},\chi\right)\;.
\end{equation}
Here, $P_{\rm g}(k,\chi)$ represents the lens 3D clustering power
spectrum, describing the fluctuations in the lens number density. We
note that for lens subsamples overlapping in radial distance, we would
expect a non-zero clustering signal when cross-correlating different
lens samples, i.e., for \mbox{$i\ne j$}.

Compiling a series of $\omega^{(ij)}(\theta_l)$ measurements within a
vector $\vec{\omega}$ of $N_\theta N_{\rm lens}(N_{\rm lens}+1)/2$
elements, we can again relate the data vector to a band power
representation of $P_{\rm g}(k,\chi)$
\begin{equation}
  \vec{\omega}=\mat{Z}\vec{f}_{\rm g}+\vec{\omega}_{\rm fid}\;,
\end{equation}
where the elements of $\vec{\omega}$ and $\mat{Z}$ are determined by
\begin{equation}
  \omega^{(ij)}(\theta)=
  \sum_{n=1}^{N_z}\sum_{m=1}^{N_k}
  Z^{(ij)}(\theta;m,n)f_{{\rm g},mn}+\omega_{\rm fid}^{(ij)}(\theta)
\end{equation}
with the basis functions
\begin{eqnarray}
  \label{eq:omegabasis}
  \lefteqn{Z^{(ij)}(\theta;m,n)=
    \frac{1}{2\pi\theta^2}\,\times}\\
  &&\nonumber
  \int_{\chi_n}^{\chi_{n+1}}\d\chi
  \frac{q^{(i)}_\chi(\chi)q^{(j)}_\chi(\chi)}{f^2_{\rm k}(\chi)}
  \int\limits_{k_mf_{\rm k}(\chi)\theta}^{k_{m+1}f_{\rm k}(\chi)\theta}\d s\,s\,
  J_0(s)\,
  P^{\rm fid}_\delta\left(\frac{s}{f_{\rm k}(\chi)\theta},\chi\right)
\end{eqnarray}
and the invariable terms, to be subtracted from the measured
correlation function
\begin{eqnarray}
  \lefteqn{\omega^{(ij)}_{\rm fid}(\theta)=
    \frac{1}{2\pi\theta^2}\,\times}\\
  &&\nonumber
  \int_{0}^{\chi_{\rm h}}\d\chi
  \frac{q^{(i)}_\chi(\chi)q^{(j)}_\chi(\chi)}{f^2_{\rm k}(\chi)}
  \int_{0}^{\infty}\d s\,s\,
  J_0(s)\,
  \cancel{P}^{\rm fid}_\delta\left(\frac{s}{f_{\rm k}(\chi)\theta},\chi\right)\;.
\end{eqnarray}
Moreover, in analogy to the previous sections, the vector
$\vec{f}_{\rm g}$ now compiles the band powers $f_{{\rm g},mn}$
relative to the fiducial matter power spectrum $P_\delta^{\rm
  fid}(k,\chi)$ in the appropriate order, and $\vec{\omega}_{\rm fid}$
is a constant offset vector that is a function of the fiducial
cosmological model filled with the values $\omega_{\rm
  fid}^{(ij)}(\theta_l)$.

By analogy with Sect. \ref{sect:pm}, a minimum variance estimator for
$\vec{f}_{\rm g}$ is
\begin{equation}
  \label{eq:omegaest}
  \vec{\hat{f}}_{\rm g}=
  \big[\mat{N}_{\rm g}^{-1}\big]^+\mat{Z}^{\rm t}\mat{N}_\omega^{-1}
  \left(\vec{\omega}-\vec{\omega}_{\rm fid}\right)\;,
\end{equation}
where 
\begin{equation}
\label{eq:noiseg}
\mat{N}_{\rm g}^{-1}=\mat{Z}^{\rm t}\mat{N}_\omega^{-1}\mat{Z}
\end{equation}
and $\mat{N}_\omega$ are the inverse estimator covariance and
measurement noise, respectively. The relative band powers $f_{{\rm
    g},mn}$ reflect the bias parameter $b^2(k,\chi)$ of the lenses
averaged over the bandwidth relative to the \emph{fiducial} matter
power spectrum.

\subsection{Mock survey}
\label{sect:survey}

\begin{figure}
  \begin{center}
    \epsfig{file=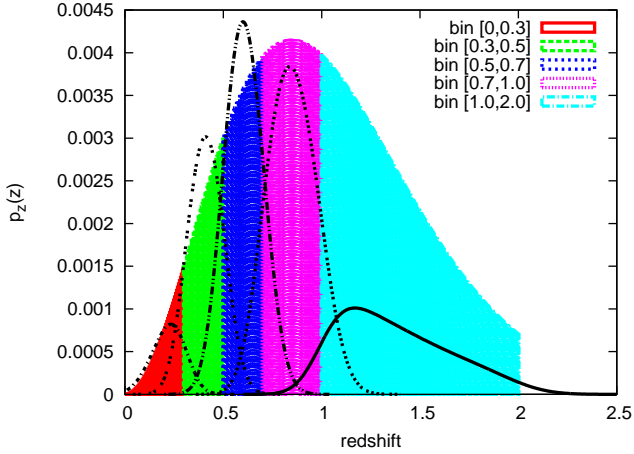,width=62mm,angle=-90}
  \end{center}
  \caption{\label{fig:pofz} Assumed redshift distribution of galaxies
    and redshift binning for the fiducial lensing survey. Shadowed
    areas highlight the sample selections in the distribution of
    redshift estimates, and the black lines show the true redshift
    distribution within the samples after including the effect of a
    redshift uncertainty of \mbox{$\sigma_z=0.05(1+z)$}. As additional
    survey parameters, we have $\bar{n}=30\,\rm arcmin^{-2}$ (sources)
    and $\bar{n}=3\,\rm arcmin^{-2}$ (lenses), an effective survey
    area of $200\,\rm deg^2$, and a shape noise variance
    $\sigma_\epsilon=0.3$. All galaxies are unbiased with respect to
    the dark matter density field.}
\end{figure}

We now move on to quantifying the accuracy of the estimators in
Eqs. \Ref{eq:xipmest}, \Ref{eq:gammatest}, and \Ref{eq:omegaest} for
prospective lensing surveys. For this purpose, we assume an effective
fiducial survey area of $200\,\rm deg^2$ in which galaxies, sources,
and lenses have a frequency distribution $p_z(z)\d z$ in redshift
given by \citep{1994MNRAS.267..323B}
\begin{equation}
  p_z(z)\propto z^\alpha\e^{-(z/z_0)^\beta}\;,
\end{equation}
where \mbox{$z_0=0.7$}, \mbox{$\alpha=2.0$}, and \mbox{$\beta=1.5$},
and the mean redshift is hence \mbox{$\bar{z}=0.9$}. To account for
the impact of redshift errors, the redshifts in this distribution are
assumed to be estimates, such as photometric redshifts based on the
photometric filter system of the survey. The estimates are unbiased
with an assumed Gaussian root mean square (r.m.s.) uncertainty of
\mbox{$\sigma_z=0.05(1+z)$}, which is similar to contemporary accuracy
levels \citep{2008A&A...480..703H}.  The total mean number density of
sources is set to \mbox{$\bar{n}=30\,\rm arcmin^{-2}$}. The intrinsic
shape noise of the sources has an r.m.s. variance of
\mbox{$\sigma_\epsilon=0.3$} for each ellipticity component. Hence,
the given survey parameters are quite optimistic in terms of the
galaxy number density and survey depth, reflecting typical parameters
of a presumably space-based survey.

Lenses constitute only ten percent of the total number of sources,
that are randomly chosen from the total source sample. The idea is
that in reality we select a subpopulation of lens galaxies to be
studied, rather than the maximum number of galaxies. In the mock data,
lenses still, however, have the same $z$-distribution as sources. In
this scenario, galaxies are unbiased with respect to the dark matter
distribution, i.e., \mbox{$b(k,\chi)=r(k,\chi)=1$} for all scales $k$
and radial distances $\chi$.

For the mock analysis, the galaxy sample is split into \mbox{$N_{\rm
    lens}=N_{\rm source}=5$} (photometric) redshift bins
$[0,0.3],[0.3,0.5],[0.5,0.7],[0.7,1],[1,2]$ with no overlap. The last
redshift bin is relatively broad, as we do not expect to obtain very
accurate redshift estimates for galaxies with $z\in[1,2]$. The
distribution inside the redshift slices is convolved with the Gaussian
redshift uncertainty $\sigma_z$ to obtain the true redshift
distributions. Owing to the uncertainty in the redshift estimators,
distributions of different samples subsequently overlap in the radial
direction. The analysis assumes that the resulting true distributions
are precisely known.  The redshift binning and distribution of
galaxies is illustrated by Fig. \ref{fig:pofz}.

The grid defining the band boundaries of the three power spectra
$P_\delta(k,\chi)$, $P_{\delta\rm g}(k,\chi)$, and $P_{\rm g}(k,\chi)$
is defined by the ranges $k\in[0.01\,h\,{\rm Mpc^{-1}},50\,h\,{\rm
  Mpc^{-1}}]$ with \mbox{$N_k=10$} log-bins and $z\in[0,2]$ for
\mbox{$N_z=5$} redshift bins. The redshift bins for the grid are
identical to the bins of the angular correlation functions. In
generally, the grid redshift limits could be chosen independently, and
the number of redshift bins might also differ. For the fiducial 3D
matter power spectrum, we adopt the fitting formula of
\citet{Smith03}. Matter and galaxies cluster exactly as in the
reference matter power spectrum, i.e., all $f_{mn}=1$.

\subsection{Measurement noise}

\begin{figure*}
  \begin{center}
    \epsfig{file=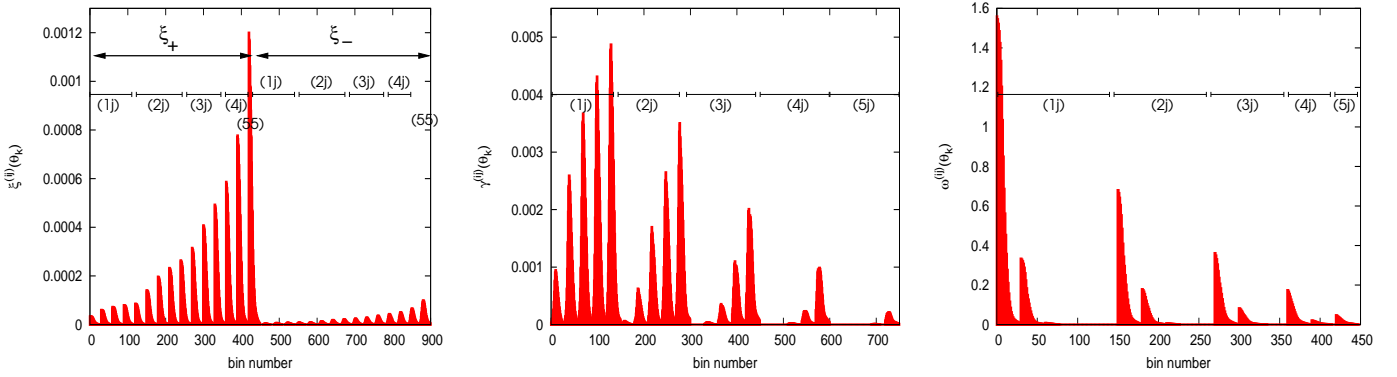,width=185mm,angle=0}
  \end{center}
  \caption{\label{fig:exdatavec} Noise-free data vectors $\vec{\xi}$
    (left), $\vec{\overline{\gamma}}_{\rm t}$ (middle) and
    $\vec{\omega}$ (right). Used are \mbox{$N_{\rm source}=5$} source
    redshift slices and \mbox{$N_{\rm lens}=5$} lens redshift slices;
    $i,j$ are indices of slices sorted by ascending mean redshift. The
    data vector elements are sorted by tomography index $(ij)$ and
    then angular separation, \mbox{$N_\theta=30$}. A spike corresponds
    to one $(ij)$ tomography block with varying $\theta$.}
\end{figure*}

The essential ingredients for quantifying the estimators' accuracy are
the noise covariances $\mat{N}_\xi$ (shear-shear correlations),
$\mat{N}_\gamma$ (galaxy-galaxy lensing), and $\mat{N}_\omega$ (lens
clustering). This includes the source shape noise, lens position
shot-noise, and cosmic variance of the actual signal. For an extensive
discussion of the noise terms in $\xi_\pm$ we refer to for example
\cite{2002A&A...396....1S}. Towards this goal, we make \mbox{$N_{\rm
    real}=4800$} realisations of mock shear and lens catalogues in
$2\times2\,\rm deg^2$ fields based on the previous survey parameters
and the matter/galaxy clustering in our fiducial cosmological model;
shear fields are Gaussian random fields and galaxy number density
fields obey log-normal statistics. The details of the mock catalogue
generation are laid out in Appendix \ref{sect:montecarlo}. Each mock
survey field is processed to estimate the cosmic shear correlations
$\xi_\pm^{(ij)}(\theta)$ \citep{2002A&A...396....1S}, the
galaxy-galaxy lensing signal $\overline{\gamma}_t^{(ij)}(\theta)$
\citep[e.g.,][]{2007A&A...461..861S} and the lens clustering
$\omega^{(ij)}(\theta)$ \citep{ls93}. The correlators are hence
estimated exactly as in a real analysis; references for the employed
estimators are given inside the previous brackets. The correlation
functions are estimated between angular separations of
\mbox{$\theta\in[3^\pprime,2\,\rm deg]$} using $N_\theta=30$
log-bins. The tomography measurements for all three correlators are
separately combined into a data vector $\vec{d}_i$ for each
realisation \mbox{$i=1\ldots N_{\rm real}$}. The data vectors have the
following structure:
\begin{itemize}
\item $\xi_\pm^{(ij)}(\theta)$: The first half of $\vec{d}_i$ contains
  estimates of $\xi_+$, whereas the second half contains the $\xi_-$
  measurements. Both the $\xi_+$ and $\xi_-$ values are sorted as
  tomography blocks of constant index $(ij)$ and ascending index $i$,
  \mbox{$i\le j$}. Within a block, values are arranged in order of
  increasing angular separation bin $\theta_l$. Since there are
  $N_\theta$ angular bins, $N_{\rm source}(N_{\rm source}+1)/2$
  tomography blocks, and two correlation functions, the data vector
  realisations $\vec{d}_i$ have in total $N_\theta N_{\rm
    source}(N_{\rm source}+1)=900$ elements.
\item $\overline{\gamma}_{\rm t}^{(ij)}(\theta)$: Likewise, GGL
  measurements are sorted in order of tomography block $(ij)$ index
  with ascending lens sample index $i$ and source sample index
  $j$. All $(ij)$ combinations are allowed, i.e., $N_{\rm lens}N_{\rm
    source}$ combinations. Inside a block, values are ordered by
  increasing lens-source angular separation. This yields in total
  $N_\theta N_{\rm lens}N_{\rm source}=750$ elements.
\item $\omega^{(ij)}(\theta)$: Measurements of the angular lens
  clustering are again sorted in order of tomography bin index $(ij)$
  with \mbox{$i\le j$} and $i,j$ being the lens sample indices; there
  are in total $N_{\rm lens}(N_{\rm lens}+1)/2$ block
  combinations. Inside a block, measurements are sorted in order of
  increasing lens-lens separation. In total, this yields $N_\theta
  N_{\rm lens}(N_{\rm lens}+1)/2=450$ elements.
\end{itemize}
Noise-free data vectors with fiducial survey parameters are depicted
in Fig. \ref{fig:exdatavec}.

As an estimator of the noise covariance based on the mock data
realisations, we devise the field-to-field variance
\begin{equation}
  \mat{C}_{\rm est}=\frac{1}{N_{\rm real}-1}\sum_{i=1}^{N_{\rm real}}
  \left(\vec{d}_i-\overline{\vec{d}}\right)
  \left(\vec{d}_i-\overline{\vec{d}}\right)^{\rm t}\;,
\end{equation}
where
\begin{equation}
  \overline{\vec{d}}=
  \frac{1}{N_{\rm real}}\sum_{i=1}^{N_{\rm real}}\vec{d}_i
\end{equation}
is the ensemble mean. The resulting matrix $\mat{C}_{\rm est}$
estimates the noise covariance $\mat{N}_\xi$, $\mat{N}_\gamma$, or
$\mat{N}_\omega$ in a $A_4=2\times2\,\rm deg^2$ survey depending on
whether $\vec{d}_i$ consists of $\xi_\pm^{(ij)}(\theta)$,
$\overline{\gamma}_{\rm t}^{(ij)}(\theta)$, or
$\omega^{(ij)}(\theta)$, respectively. We scale this to the expected
noise in a larger $A_{200}=200\,\rm deg^2$ survey by multiplying
$\mat{C}_{\rm est}$ with \mbox{$A_4/A_{200}=1/50$}. This mimics a
survey of area $A_{200}$ consisting of 50 statistically independent
patches of size $A_4$ each. Finally, since the band power estimators
require the \emph{inverse} noise covariance, we utilise the inverse
covariance estimator of \citet{2007A&A...464..399H}, which is based
upon $\mat{C}^{-1}_{\rm est}$.

For the covariance estimation and the mock analysis, lenses are a
randomly chosen subsample of the sources (ten percent).  In
particular, sources are clustered as the projected dark matter
distribution up to the level of the Poisson shot noise. The choice of
identical redshift bins for lenses and sources is due purely to the
means by which the mock survey galaxy catalogues are generated. In the
estimator formalism, lenses and sources may have different radial
distributions and the number of subsamples may differ as well. The
$z$-slices in the mock data overlap due to the adopted redshift
uncertainty. Therefore, we have a non-vanishing
$\omega^{(ij)}(\theta)$ for \mbox{$i\ne j$}, which can be seen in the
right panel of Fig. \ref{fig:exdatavec}.

\begin{figure*}
  \begin{center}
    \epsfig{file=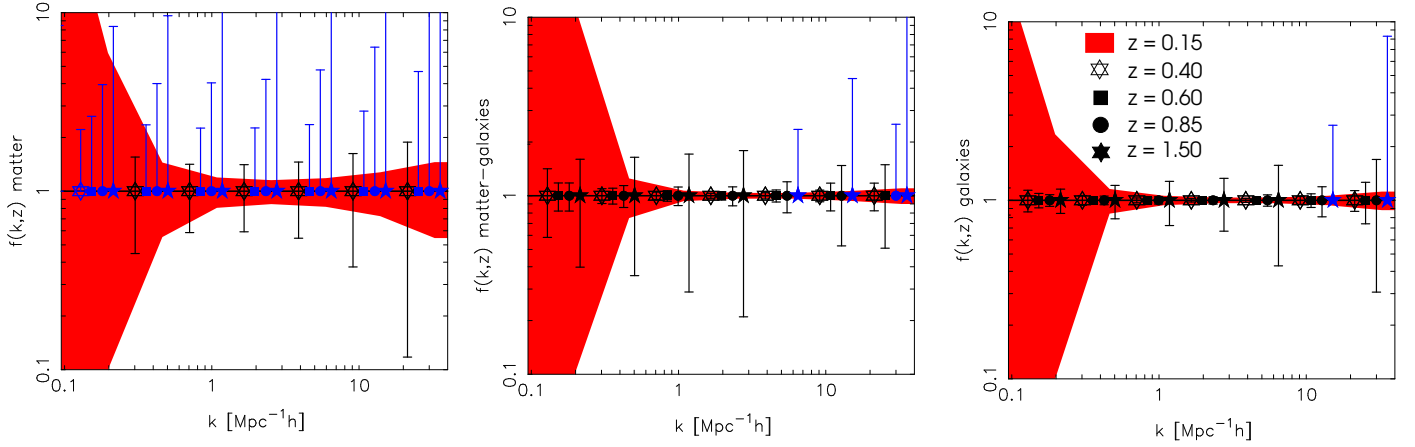,width=185mm,angle=0}
  \end{center}
  \caption{\label{fig:mvestimate} A simulated reconstruction of the 3D
    band matter power spectrum $P_\delta(k,\chi)$ (left),
    galaxy-matter cross-correlation power spectrum $P_{\delta\rm
      g}(k,\chi)$ (middle), and galaxy power spectrum $P_{\rm
      g}(k,\chi)$ (right) utilising the minimum variance estimators in
    Sect. \ref{sect:quadratic}. Shown are the average estimated band
    powers $f_{mn}$ relative to the fiducial band power spectrum. The
    shadowed area in the background indicates the $68\%$ confidence
    region of the lowest redshift bin \mbox{$\bar{z}=0.15$}.  The data
    points denote estimates at higher redshifts. The error regions
    indicate the $1\sigma$ uncertainties of the estimators via
    Eq. \Ref{eq:noisedelta}, \Ref{eq:noisedeltag}, or
    \Ref{eq:noiseg}. Data points in blue colour or with asymmetric
    error bars indicate estimates that are consistent with zero. Note
    that data points within the same $k$-bin are slightly shifted with
    respect to each other along the $k$-axis for clarity.}
\end{figure*}

\subsection{Forecasts}
\label{sect:mvforecast}

Using the fiducial survey and the predicted measurement noise,
constraints on the three power spectra are shown in
Fig. \ref{fig:mvestimate}. The figures are obtained by applying the
estimators to a noise-free data vector (Fig. \ref{fig:exdatavec}),
whereas the error bars are based on estimates of the measurement noise
in the mock survey, that is either Eq. \Ref{eq:noisedelta},
\Ref{eq:noisedeltag}, or \Ref{eq:noiseg}. The data points or the
centre of the shadowed region therefore simulate the statistical
average of a band power estimation process for an infinite number of
similar surveys. The error bars or regions, however, are the predicted
$1\sigma$ uncertainties in a typical single $200\,\rm deg^2$ survey.
As can be seen, the average data points lie exactly on \mbox{$f_{\rm
    mn}=1$}, indicating that they were evaluated with an unbiased
estimator.

In general, the constraints involving cosmic shear become worse with
increasing redshift owing to the decreasing number of background
sources at higher redshifts. For the matter power $f_{\delta,mn}$, we
find only upper limits in the redshift bins \mbox{$z>0.5$} or for
scales \mbox{$k\lesssim0.3\,h\rm Mpc^{-1}$}, but reasonable
constraints for \mbox{$z\lesssim0.4$} and \mbox{$0.5\le k\,h^{-1}{\rm
    Mpc}\le10$}. A similar conclusion can be drawn for $f_{{\delta\rm
    g},mn}$, albeit the constraints are here considerably improved,
especially at higher redshift. Under our assumptions, the galaxy band
power spectrum of the lenses, $f_{{\rm g},mn}$, is extremely
well-recovered. The detection fidelity depends mainly on the number of
lenses within a redshift bin.  In reality, this number is likely to be
even smaller as in the mock survey, depending on the galaxy population
selected for investigation \citep[see, e.g.,][]{2011ApJ...736...59Z}.

Since the correlation functions cover only a certain angular range, we
may anticipate obtaining biased estimates for large-scale modes
$k$. This bias can be quantified by utilising both
Eq. \Ref{eq:biasest} for $\vec{f}^{\rm ref}_\delta$ and equivalent
bias indicators for the two other band powers
$\hat{\vec{f}}_{\delta\rm g}$ and $\hat{\vec{f}}_{\rm g}$. For the
band limits and fiducial power spectrum chosen, the bias is smaller
than 0.1 percent and therefore no concern to us here. We note,
however, that a bias clearly becomes visible for \mbox{$P^{\rm
    fid}_\delta(k,\chi)\equiv1$} and \mbox{$k\lesssim0.2\,h\rm
  Mpc^{-1}$}. Therefore, a realistic matter power spectrum when used
as a reference aids the stability of the minimum variance estimators.

\begin{figure*}
  \begin{center}
    \epsfig{file=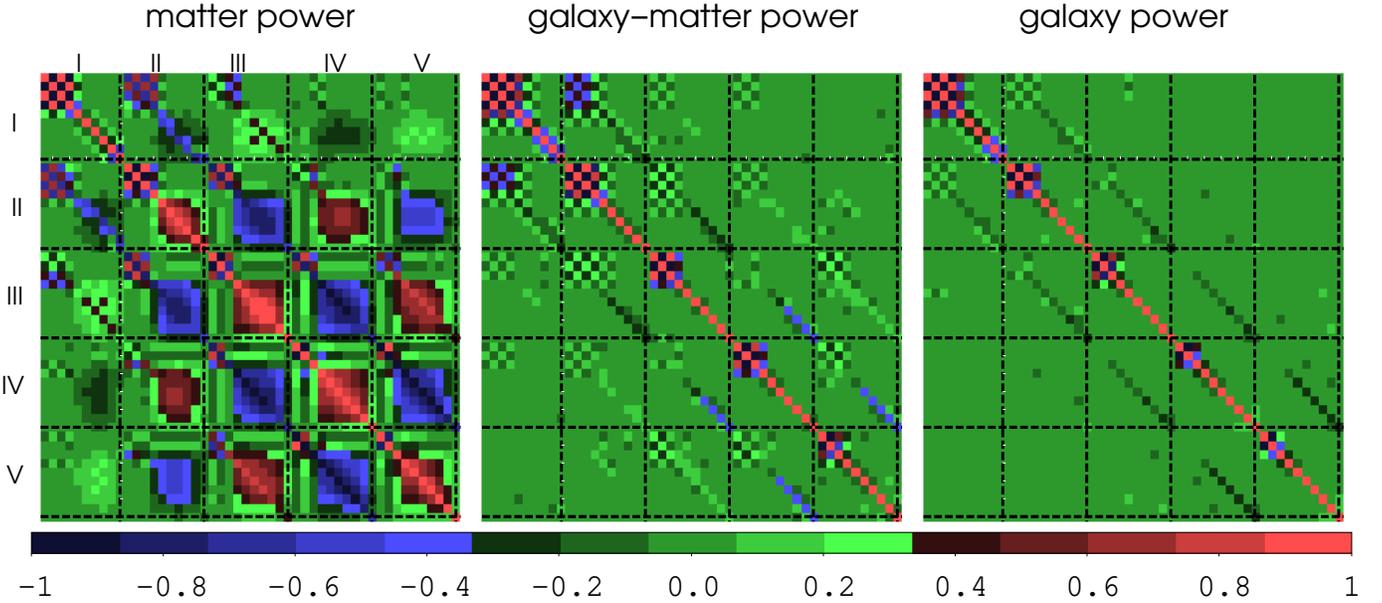,width=185mm,angle=0}
   \end{center}
   \caption{\label{fig:mvcorr} Correlation matrix of
     uncertainties in the relative band powers (minimum variance
     estimators). Roman numbers indicate the redshift bins
     ($\bar{z}=0.15,0.4,0.6,0.85,1.5$ for I-V, respectively). Within
     blocks, highlighted by dashed lines, the correlation between
     $k$-bins of either identical $z$-bins (diagonal blocks) or
     different $z$-bins (off-diagonal blocks) is plotted. The mean
     wave numbers of the $k$-bins are
     $\bar{k}=0.01,0.02,0.05,0.13,0.30,0.71,1.66,3.88,9.10,21.33$ in
     units of $h\,\rm Mpc^{-1}$.}
\end{figure*}

The correlation of the estimator uncertainties are depicted in
Fig. \ref{fig:mvcorr}. Uncertainties in the estimates of the relative
band powers $f_{\delta,mn}$ clearly become increasingly correlated at
higher redshift. Most independent estimates are located in the lowest
redshift bin, where the constraints are also tightest
(Fig. \ref{fig:mvestimate}). For \mbox{$z>0.4$}, the errors evidently
become strongly correlated because all shear-shear correlation
functions are sensitive to the same matter density fluctuations in
front of the sources: For constant $k$, $f_{\delta,mn}$ adjacent in
redshift are anti-correlated. For neighbouring bands in the same
redshift bin, errors are also strongly correlated for
\mbox{$z>0.4$}. Therefore, there is little statistically independent
information in the $f_{\delta,mn}$ at higher redshift. The correlation
of errors becomes small for the galaxy-matter band power
$f_{{\delta\rm g},mn}$ and is almost absent for the galaxy band power
$f_{{\rm g},mn}$. However, all estimates still exhibit strong
correlations between adjacent $k$-bins in the large-scale regime
(small $k$) because the bins are associated with density modes that
are barely touched by the angular correlation functions of limited
angular range.

\section{Bayesian analysis}
\label{sect:bayes}

The minimum variance estimators of the foregoing section are quick and
easy to apply to the data. Moreover, their statistical errors are
equally straightforward to assess. However, they have the disadvantage
that band power estimates can also become negative for the
auto-correlation power spectra $P_\delta(k,\chi)$ and $P_{\rm
  g}(k,\chi)$, which by their definitions is not allowed.  One
frequently finds band power estimates oscillating about zero where
errors are large. Therefore, the question arises of whether one can
refine the analysis to explicitly incorporate non-negative
powers. This refinement may also help to improve the recovery of the
matter power spectrum, which in the foregoing minimum variance
analysis was only successful at low redshifts, where the majority of
band power factors $f_{\delta,mn}$ had only upper limits. We note that
the cross-correlation power $P_{\delta\rm g}(k,\chi)$ can in principle
become negative, if the galaxy correlation parameter $r(k,\chi)$ is
negative.

\subsection{Method}

A refinement can be achieved in the framework of a Bayesian
analysis \citep[cf.][]{2003Book...MACKAY}, where we determine the
posterior likelihood
\begin{equation}
  p\big(\vec{m}|\vec{d}\big)\propto
    {\cal L}\big(\vec{d},\vec{m}\big)\times 
    p_{\rm prior}\big(\vec{m}\big)
\end{equation}
of our model parameters $\vec{m}$ -- for instance, the band power
values $\vec{f}_\delta$ in the case of $P_\delta(k,\chi)$ -- for the
observed tomography correlation functions $\vec{d}$, which is
$\vec{\xi}$ for the example mentioned; ${\cal L}(\vec{d},\vec{m})$ is
the data likelihood function and $p_{\rm prior}(\vec{m})$ the a-priori
information on the band power (prior).

In the case of $\vec{f}_\delta$, we solely assume a prior that
enforces positive or zero band powers
\begin{equation}
  p_{\rm prior}\big(\vec{f}_\delta\big)=
  \prod_{n=1}^{N_z}\prod_{m=1}^{N_k}H\big(f_{\delta,mn}\big)\;,
\end{equation}
where $H(x)$ is the Heaviside function of $x$. A conceivable
modification of the prior could consist of including available
empirical information on the matter power spectrum. For the likelihood
function, we presume Gaussian noise
\begin{equation}
  \label{eq:likelihood}
  {\cal L}_\xi\big(\vec{\xi},\vec{f}_\delta\big)\propto
    \exp{\left(-\frac{1}{2}\left[\vec{\xi}-\mat{X}\vec{f}_\delta-\vec{\xi}_{\rm fid}\right]^{\rm t}\mat{N}_\xi^{-1}\left[\vec{\xi}-\mat{X}\vec{f}_\delta-\vec{\xi}_{\rm fid}\right]\right)}\;,
\end{equation}
with a constant noise covariance $\mat{N}_\xi$ that does not depend on
the matter power spectrum. The likelihood functions ${\cal L}_\gamma$
and ${\cal L}_\omega$ for $\vec{f}_{\delta\rm g}$ and $\vec{f}_{\rm
  g}$, respectively, are defined accordingly.  Furthermore, for
$f_{{\rm g},mn}$ we employ a prior similar to the $f_{\delta,mn}$
prior, but allow $f_{{\delta\rm g},mn}$ to be negative (negative
correlations).

Another convenience of the Bayesian approach is that one can easily
accommodate a transformation of model parameters in the analysis,
yielding the statistics of an alternative parameter set. We take
advantage of this by also determining the posterior likelihoods of
band galaxy biasing parameters $(b_{mn},r_{mn})$ by expressing
$\vec{f}_{\delta\rm g}$ and $\vec{f}_{\rm g}$ as
\begin{eqnarray}
  \label{eq:bias}
  f_{{\delta\rm g},mn}&=&r_{mn}b_{mn}\,f_{\delta,mn}\;,\\
  \nonumber
  f_{{\rm g},mn}&=&b^2_{mn}\,f_{\delta,mn}\;,
\end{eqnarray}
for all $m=1\ldots N_k$ and $n=1\ldots N_z$. Here the prior ensures
that the bias factors $b_{mn}$ and the relative band powers
$f_{\delta,m}$ are positive, thereby directly constraining $b(k,\chi)$
and $r(k,\chi)$, provided that we employ the same reference fiducial
power spectrum for the matter, galaxy-matter, and galaxy power
spectrum. In this variant, we combine the information on all
$(f_{\delta,mn},f_{{\delta\rm g},mn},f_{{\rm g},mn})$ to constrain the
biasing parameters. Consequently, we now have a combined posterior
likelihood instead of the three separate ones in the previous case
\begin{eqnarray}
  \lefteqn{p\big(\vec{f}_\delta,\vec{b},\vec{r}|\vec{d}\big)\propto
    p_{\rm prior}\big(\vec{r},\vec{b},\vec{f}_\delta\big)\,\times}
  \\\nonumber
  &&{\cal L}_\xi\big(\vec{\xi},\vec{f}_\delta\big)
  {\cal L}_\gamma\big(\vec{\overline{\gamma}}_{\rm
        t},\vec{f}_{\delta\rm g}=\vec{br}\vec{f}_\delta\big)
  {\cal
    L}_\omega\big(\vec{\omega},\vec{f}_{\rm
      g}=\vec{b}^2\vec{f}_\delta\big)\;.   
\end{eqnarray}
We assume that noise correlations between $\xi_\pm$,
$\overline{\gamma}_{\rm t}$, and $\omega$ estimators are negligible,
hence likelihood function factors are combined as shown, where
$(\vec{b},\vec{r})$ denotes the biasing parameters and
$(\vec{brf}_\delta,\vec{b^2f}_\delta)$ provide an abbreviated version
of Eq. \Ref{eq:bias}.

We sample the posterior likelihoods numerically by devising the Markov
chain Monte Carlo (MCMC) technique \citep[e.g.,][]{2003Book...MACKAY},
where the noise covariances of the minimum variance estimators
$\hat{\vec{f}}$ are used as yardsticks to set up the proposal function
for the Metropolis-Hastings algorithm.

\begin{figure*}
  \begin{center}
    \epsfig{file=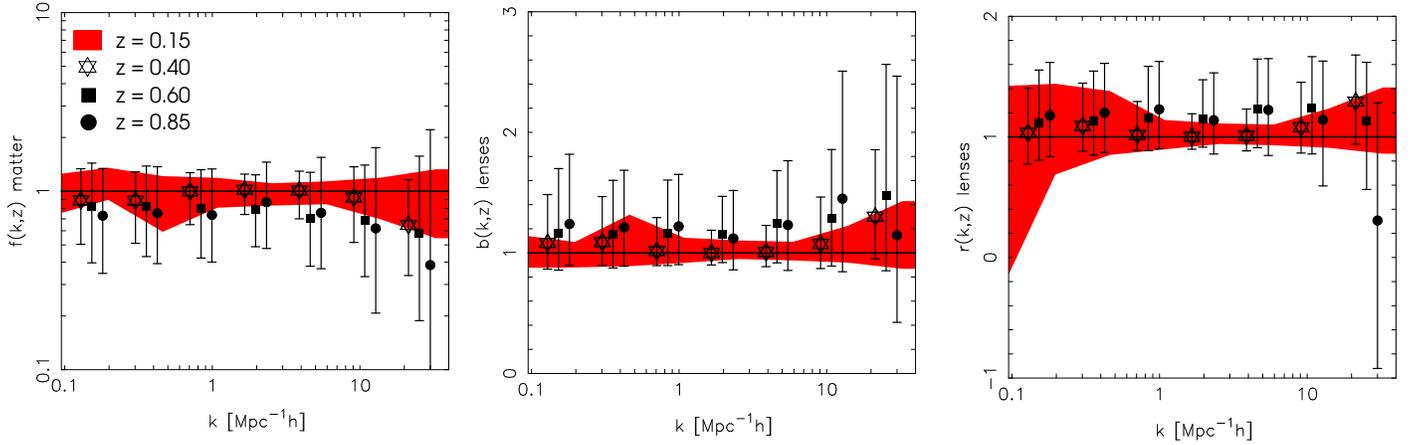,width=185mm,angle=0}
  \end{center}
  \caption{\label{fig:bayeslogflat} Bayesian MCMC analysis of the
    fiducial survey estimating the matter band power (left; relative
    to the fiducial power spectrum) and the galaxy biasing parameters
    (middle: bias factor; right: correlation factor). The shadowed
    area in the background highlights a $68\%$ credibility region
    about the marginalised posterior mean for the lowest redshift bin
    \mbox{$\bar{z}=0.15$}, whereas data points delineate the $68\%$
    credibility regions about the mean at higher redshifts. The
    highest, most poorly constrained redshift bin is not included
    here.  The results are subject to some numerical noise caused by
    the adopted MCMC technique.}
\end{figure*}

\begin{figure*}
  \begin{center}
    \epsfig{file=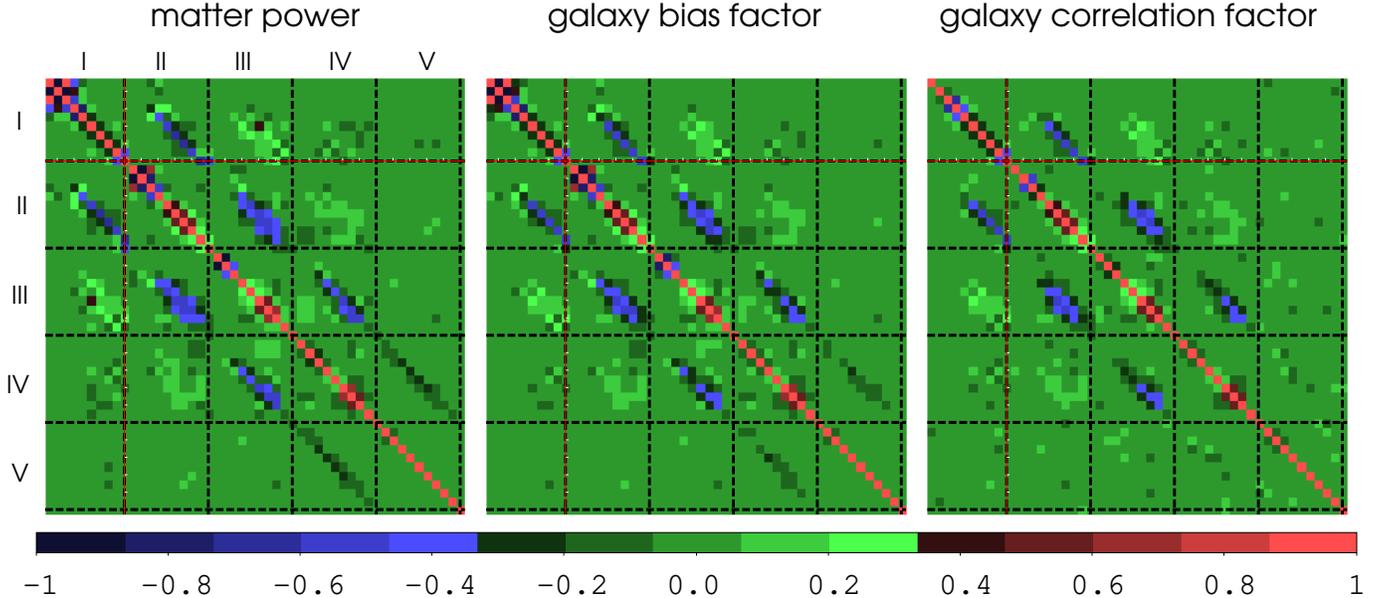,width=185mm,angle=0}
   \end{center}
   \caption{\label{fig:bayescorr} Correlation matrix of band power and
     galaxy biasing parameter estimates in the Bayesian framework (see
     Fig. \ref{fig:bayeslogflat}). See caption of
     Fig. \ref{fig:mvcorr} for more information.}
\end{figure*}

\subsection{Forecasts}
\label{sect:bayesforecast}

Fig. \ref{fig:bayeslogflat} shows the constraints on the fiducial
survey (Sect \ref{sect:survey}) obtained via the Bayesian analysis. We
show only the variant that uses biasing parameters to express the
galaxy-matter and galaxy power spectrum. As before with the minimum
variance estimators, the constraints are obtained by using a
noise-free data vector as input data, as shown in
Fig. \ref{fig:exdatavec}. The likelihood functions ${\cal
  L}_{\xi,\gamma,\omega}$ assume a level of noise that is similar to
that in the $200\,\rm deg^2$ fiducial survey. This combination
basically simulates the typical posterior of a $200\,\rm deg^2$
survey. We plot the means of the marginalised one-dimensional (1D)
posteriors of $(f_{\delta,mn},b_{mn},r_{mn})$ and the r.m.s. variance
in their MCMC values above (upper error bars) or below the mean (lower
error bars). This defines a 68 percent credibility region for every
parameter.  The prior is defined to constrain the galaxy biasing
parameters a priori for \mbox{$b(k,\chi)\in[0,4]$} and
\mbox{$r(k,\chi)\in[-2,2]$} (flat priors). These priors broadly
delineate the regime that is expected for galaxy bias
\citep[e.g.][]{2001MNRAS.321..439G,2005PhRvD..71d3511S}.

We note that owing to the subtraction of Poisson shot noise in the
definition of $P_{\rm g}(k,\chi)$ (Eq. \ref{eq:galpower}), the modulus
of the correlation coefficient $r$ can be larger than unity. A value
of \mbox{$|r|>1$} indicates a discrete galaxy sampling that differs
from a Poisson process as, for instance, predicted on small scales in
the halo model \citep[cf.][]{2001MNRAS.321..439G}. The Monte Carlo
mock data employed in this study uses an explicit Poisson process to
generate galaxy positions out of a smooth number density field
(Appendix). 

For $\vec{f}_\delta$, a flat prior on a logarithmic-scale was
used. However, we found that a prior flat on a linear scale yields
similar results for $\vec{f}_\delta$ in our case. This result, which
is based on a noise-free data vector, shows that the mean of the 1D
posterior probability distribution function (p.d.f.) is \emph{not} an
unbiased estimator of any $f_{mn}$, but the mean is biased to too low
values for $f_{\delta,mn}$ and biased too high for $b_{mn}$ and for
most factors $r_{mn}$. This result also depends on the adopted priors,
particularly when the constraints from the data are weak. For example,
the flat prior for $b_{mn}$ tends to predict values of $b_{mn}$ of
around two. We note, however, that here the true \mbox{$f_{mn}=1$}
always falls into the above defined 68 percent credibility region such
that the bias appears to be at least smaller than the
r.m.s. uncertainty caused by measurement noise. This can be explained
in the following way. First, the posterior mean $\ave{f_{mn}}$ is an
optimal estimator $f^{\rm est}_{mn}$ in the sense that it minimises
the average estimator error
\begin{equation}
  \label{eq:esterror}
  \Ave{\Delta^2 f_{mn}^{\rm est}}:=
  \int\d f_{mn}\,p\big(f_{mn}|\vec{d}\big)
  \,\left(f_{mn}^{\rm est}-f_{mn}\right)^2
\end{equation}
as to the true $f_{mn}$ with posterior $p(f_{mn}|\vec{d})$ because
\begin{equation}
  \frac{\d \ave{\Delta^2 f_{mn}^{\rm est}}}{\d f^{\rm
      est}_{mn}}=0~{\rm for}~
  f^{\rm est}_{mn}\equiv\ave{f_{mn}}=\int\d f_{mn}\,p(f_{mn}|\vec{d})\,f_{mn}\;.
\end{equation}
Second, the average error when adopting \mbox{$f^{\rm
    est}_{mn}=\ave{f_{mn}}$} is simply the r.m.s. variance in the
posterior, as can be seen by substituting $\ave{f_{mn}}$ for $f^{\rm
  est}_{mn}$ in Eq. \Ref{eq:esterror}.

As with the minimum variance estimators, the strongest constraints are
found for the regime \mbox{$0.5\lesssim k\,h^{-1}\,{\rm
    Mpc}\lesssim10$}. The constraints on the band matter power
spectrum, for which we had upper limits only for the minimum variance
estimates, are clearly improved by the assertion of positive
auto-power spectra. We note the apparent strong improvement for
\mbox{$k\lesssim0.5\,h\rm Mpc^{-1}$}, which is merely a combined
effect of the priors on $f_{\delta, mn}$ and the bias factor and,
possibly, also the information on the galaxy clustering on large
scales from the data.

In addition, the MCMC results are devised to estimate the correlation
of the errors as a correlation matrix in
Fig. \ref{fig:bayescorr}. This also shows an improvement for
$f_{\delta,mn}$. The correlation of errors is softened, hence much
less pronounced than in Fig. \ref{fig:mvcorr}, although we can still
make out the anti-correlation among the bins that are adjacent in
redshift. The (band) galaxy bias and correlation factors $b_{mn}$ and
$r_{mn}$ have errors that are very similar to those of $f_{\delta,mn}$
since the correlation pattern of the latter is imprinted in the galaxy
biasing parameters owing to the mixing of $f_{\delta,mn}$,
$f_{{\delta\rm g},mn}$, and $f_{{\rm g},mn}$ within the Bayesian
analysis.

\section{Discussion and conclusions}
\label{sect:discuss}

This paper has presented methods for constraining from tomography data
the deprojected power spectra of the matter clustering and the related
clustering of galaxies. Moreover, a fiducial survey has been employed
to determine the prospects for the application of this method to real
data.

The prerequisite for a successful application of this technique is the
availability of radial (redshift) information for survey galaxies,
which will be provided by ongoing lensing surveys and surveys in the
near future. On the basis of this radial information, sources and
lenses can be divided into subsamples with known radial distributions,
which are disjunct slices in the optimal case, thereby probing matter
and galaxy clustering at different effective redshifts and on
different physical scales.  From the cosmic shear tomography alone,
the spatial matter power spectrum can be measured, which represents an
advance on older lensing methodologies
\citep{2002A&A...396....1S,2003MNRAS.346..994P,2005MNRAS.363..723B}.
Augmenting this with measurements of galaxy-galaxy lensing and angular
lens clustering for all pair combinations of galaxy subsamples,
constraints on the galaxy biasing can also be made as a function of
redshift and length-scale. This improves older non-parametric lensing
techniques for probing galaxy biasing
\citep{1998ApJ...498...43S,1998A&A...334....1V} since tomography
correlations can be directly traced back to the underlying 3D power
spectrum. In particular, the calibrations of bias parameters due to
projection effects developed by \citet{2002ApJ...577..604H} are no
longer required. This promises to provide a simplified comparison to
theoretical galaxy or dark matter models
\citep[cf.][]{2001ApJ...558..520Y,2004ApJ...601....1W,2005Natur.435..629S}.

\subsection{Methodology}

We have shown in this paper that by approximating the 3D power spectra
by band power spectra relative to a fiducial band power spectrum,
there is a straightforward linear relation between the observable
angular correlation functions (shear-shear correlations, mean
tangential shear about lenses, and lens clustering) and the 3D power
spectra (matter power, matter-galaxy cross power, and galaxy power),
where Limber's approximation applies. This study proposed two
approaches to inverting this linear relation in a statistical way:
\begin{itemize}
\item The first approach makes use of minimum variance estimators that
  provide estimates of the relative 3D band powers. The approach is
  similar to that of \citet{2003MNRAS.346..994P}, but now permits the
  full amount of tomography information to be used. By doing so we can
  now, therefore, observe the time-evolving power spectra rather than
  the average power at an effective depth of the survey. Furthermore,
  by choosing a theoretical power spectrum as reference, the minimum
  variance estimators allow us to compare a theoretical band power
  directly to the data. Differences between the theory and data are
  thereby revealed by a relative band power significantly
  \mbox{$f_{\rm mn}\ne1$}.
\item The second approach is a full Bayesian analysis for which we
  can, in addition, express the galaxy-matter and galaxy power
  spectrum in terms of the linear galaxy biasing functions $b(k,\chi)$
  and $r(k,\chi)$. The biasing functions are related to the angular
  correlation functions in a nonlinear fashion. The Bayesian framework
  automatically accounts for this nonlinearity in the posterior
  likelihood of the biasing parameters, being the output of the
  analysis. Moreover, we can easily implement the prior knowledge that
  the matter and galaxy power, by definition variances, have to be
  positive or zero.  On the other hand, the disadvantage of the
  Bayesian approach is that we have to specify the complete noise
  statistics of the measurements, whereas the minimum variance
  estimators formally only require the noise covariances. For the
  scope of this paper, we adopted a multivariate Gaussian noise
  model. This does not exactly hold true
  \citep{2009A&A...504..705S,2011arXiv1105.3672K}, but a more
  sophisticated noise model can be built into the analysis by
  modifying the likelihood function in Eq. \Ref{eq:likelihood} in
  future applications. Moreover, the Bayesian framework also allows us
  to add prior information about the matter power or galaxy biasing
  from other observations.
\end{itemize}

Both approaches rely on a known angular diameter distance, $f_{\rm
  k}(\chi)$, in addition to a specified relation between radial
comoving distance $\chi$ and redshift for a given fiducial cosmology
model. Although this may sound like a strong restriction, those
geometric properties of the Universe are known to great accuracy and
are consistently inferred by a wide range of observations
\citep[e.g.,][]{2011ApJS..192...18K}. The nonlinear structure of
matter at \mbox{$z\lesssim1$}, on the other hand, is prone to a larger
uncertainty as discussed, for instance, by
\citet{2011MNRAS.415.3649V}.

\subsection{Mock survey} 

To estimate the performance of the estimators, we assumed a fiducial
survey with a $200\,\rm deg^2$ area, a mean galaxy redshift of
$\bar{z}=0.9$, reaching as deep as $z=2.0$ for usable redshift
estimates, and a source number density of \mbox{$\bar{n}=30\,\rm
  arcmin^{-2}$} (Sect. \ref{sect:survey} for more details). Lenses are
a subsample of ten percent of the source catalogue, but have an
identical $z$-distribution.  This is because a smaller galaxy
population is usually studied as lenses. We assumed that the angular
correlation functions can safely be measured between 3 arcsec and $2$
degree. Separations much smaller than 3 arcsec will be difficult to
measure owing to the apparent size of smeared galaxy images or the
pixel size in the instruments. Separations larger than $\sim2$ degree
are principally possible, but this would extend into a regime where
the flat-sky approximation, used for the formalism of this paper,
would begin to fail. This may be remedied by a full-sky treatment
comparable to \citet{2005PhRvD..72b3516C}.

The fiducial parameters were chosen to be somewhere in-between typical
figures of current or near future surveys. The assumed survey depth
is, however, probably more on the side of deeper space-based lensing
surveys such as
COSMOS\footnote{\url{http://cosmos.astro.caltech.edu/astronomer/hst.html}}
or the upcoming Euclid
mission\footnote{\url{http://sci.esa.int/euclid}, see also
  \citet{2011arXiv1110.3193L}}, rather than on the side of shallower
ground-based surveys such as CFHTLS, KiDS, or Pan-STARRS. On the other
hand, the adopted number density of sources is smaller than expected
for a spaced-based mission but still higher than the typical number
density achieved in ground-based surveys (\mbox{$\sim10\,\rm
  arcmin^{-2}$}).  Moreover, the modest fiducial survey area of
$200\,\rm deg^2$ reflects the order-of-magnitude for a contemporary
ground-based lensing survey, that is capable of delivering cosmic
shear data such as the CFHTLS. In the not too distant future, this
figure is likely to be increased by another order of magnitude.

\subsection{Simulated noise covariance}

The predicted noise covariance of the correlation functions was
derived from realisations of mock surveys (Appendix
\ref{sect:montecarlo}). In analogy to the Monte Carlo method put
forward in \citet{2004A&A...417..873S}, I included a new feature to
account for the clustering of galaxies, obeying a log-normal
statistics, and to include a mock galaxy-galaxy lensing signal. This
approach is imperfect in several respects:
\begin{itemize}

\item The assumption of Gaussian statistics for the cosmic shear
  fields is inaccurate for nonlinear scales (roughly ten arcmin or
  smaller), which leads to an underestimation of the shear noise
  covariance
  \citep{2007MNRAS.375L...6S,2009A&A...504..689H,2011arXiv1105.3980H}.
  The adopted log-normal statistics \citep{1993ApJ...417...36B} of the
  galaxy clustering, on the other hand, is presumably a good
  approximation of the galaxy-galaxy lensing and galaxy clustering
  covariance.

\item As realisations on grids are performed (grid sizes are
  $4\times4\,\rm deg^2$ from which the smaller patches are cropped),
  we miss angular modes on scales larger than $\gtrsim4\,\rm deg$
  \citep{2004A&A...417..873S}. Computation time constraints in our
  approach do not allow much larger fields with more galaxies.

\item The covariance is estimated on a patch-by-patch basis, which
  assumes a discontinuous, patchy survey rather than a contiguous
  survey area where galaxy pairs across patches could also be used to
  estimate the angular correlation functions. This reduces the
  effective number of galaxy pairs and yields more noise than for a
  realistic contiguous survey area.

\end{itemize}
Although the last effect presumably partly compensates for the two
others, we should expect an overly optimistic noise covariance and
hence possibly overly optimistic constraints on the band power
spectra.  In the future, a more accurate estimate could be obtained by
ray-tracing through cosmological structure formation simulations
\citep{2000ApJ...537....1W,2009A&A...499...31H,2011MNRAS.414.2235K,2011MNRAS.415..881L}. As
the noise covariances for the tomography correlators are necessarily
large in size (Fig. \ref{fig:exdatavec} for bin numbers), however, a
large number of statistically independent realisations would be
required (this paper uses 4800 realisations, thus in total $19200\,\rm
deg^2$) to properly estimate the inverse noise covariance
\citep{2007A&A...464..399H}.

\begin{figure}
  \begin{center}
    \epsfig{file=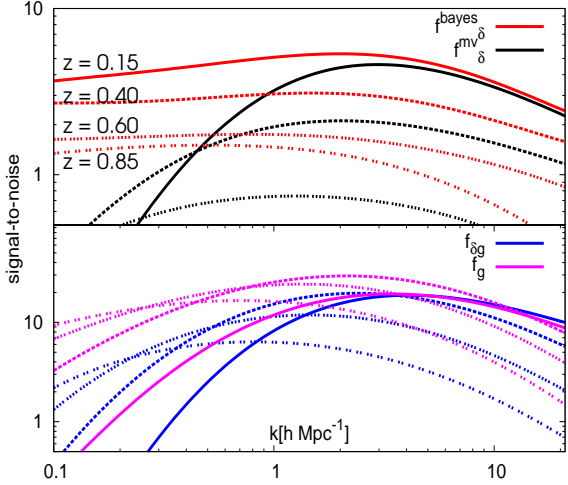,width=85mm,angle=0}
   \end{center}
   \caption{\label{fig:sn} Predicted signal-to-noise ratio of the
     relative band powers $f_{mn}$ for different redshift bins (see
     key for line styles in the top panel). Minimum variance
     estimators and Bayesian analysis give comparable results except
     for the matter band power $f_{\delta,mn}$ (top panel), which here
     combines constraints from lensing and galaxy clustering with
     priors \mbox{$b\le4$} and \mbox{$|r|\le2$} in the Bayesian
     approach. The bottom panels shows the signal-to-noise of the
     minimum variance estimates.}
\end{figure}

\begin{figure}
  \begin{center}
    \epsfig{file=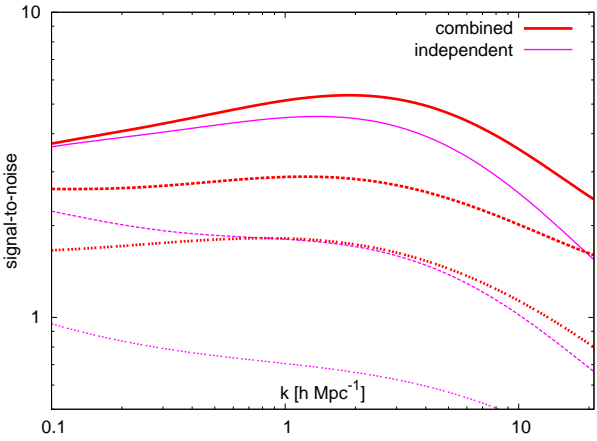,width=60mm,angle=-90}
   \end{center}
   \caption{\label{fig:sn3} Signal-to-noise ratio per band of the
     reconstructed matter power spectrum $f_{\delta,mn}$ in a Bayesian
     analysis, either for an analysis of the shear tomography alone
     (``independent''), or when combining galaxy clustering, lensing
     with priors \mbox{$b\le4$}, \mbox{$|r|\le2$}
     (``combined''). Plotted are the three first redshift bins
     $\bar{z}=0.15,0.4,0.6$ (top to bottom line).}
\end{figure}

\subsection{Reconstructed band powers}

The fiducial survey predictions are summarised by
Figs. \ref{fig:mvestimate} and \ref{fig:bayeslogflat} for the minimum
variance and Bayesian approach, respectively. The latter does not
include the highest redshift bin, \mbox{$\bar{z}=1.5$}, because of its
poor constraints.  Figs. \ref{fig:mvcorr} and \ref{fig:bayescorr}
visualise the correlation of errors in the estimates.  The correlation
of errors is quite strong for $f_{\delta,mn}$, but considerably weaker
for $f_{{\delta\rm g},mn}$ and $f_{{\rm g},mn}$ (similar between the
minimum variance and Bayesian technique). Compared to the minimum
variance estimator, the Bayesian analysis eases the strong correlation
of errors in the matter band power spectrum,.

We adopted \mbox{$N_z=5$} redshift bins and \mbox{$N_k=10$} k-bins on
a logarithmic-scale. The MCMC algorithm is well-suited to fitting many
free model parameters simultaneously (here $N_kN_z=50$), but converges
slower if their number becomes larger. For this reason, a small number
of bins is desirable.  Because adjacent bands are already moderately
to strongly correlated in the $k$- or $z$-direction, especially for
$f_{\delta,mn}$, a much finer binning than the one adopted is probably
not necessary, although this should be quantified in an objective way
for future applications.  A principal component analysis
\citep[e.g.,][]{1997ApJ...480...22T}, based on the minimum variance
estimator noise, could be employed to identify the most significant
modes in the $(k,z)$-plane and find an optimised representation of the
power spectra for the Bayesian analysis. Two interesting binning
extremes are conceivable:
\begin{itemize}
\item On the one hand, we could use \mbox{$N_k=1$} and a larger number
  of redshift bins. Using a fiducial 3D power spectrum with a
  reasonably realistic shape but normalised at all redshifts to the
  variance in the local Universe, we could focus on the overall growth
  of the fluctuations from the tomography data alone. This would
  result in an analysis similar to the one in
  \citet{2005MNRAS.363..723B}.
\item On the other hand, we could use \mbox{$N_z=1$} and a larger
  number of $k$-bins. If the time evolution of the structure growth is
  realistically built into the fiducial power spectrum, then the
  methodology from this paper would infer the deviation from the
  fiducial power as a function of scale averaged over the full
  redshift range. This would focus on deviations from the shape of the
  theoretical power spectrum, taking the expected structure growth out
  of the equation.
\end{itemize}

As long as estimates are very noisy -- auto-correlation band powers
are consistent with zero in the minimum variance approach -- the
Bayesian approach provides tighter constraints
(Fig. \ref{fig:sn}). Both approaches are otherwise comparable.  The
signal-to-noise ratio in the Bayesian analysis equals the mean of the
marginalised posterior divided by the variance about the mean.  On
large scales $k\lesssim1\,h\rm Mpc^{-1}$, however, the signal-to-noise
does not fall off as quickly as for the minimum variance estimators:
it forms instead a plateau. By performing a Bayesian analysis for the
matter power spectrum combined with galaxy clustering and one analysis
based on shear tomography alone, we have found that the plateau is the
result of the asserted positive power $P_\delta(k,\chi)\ge0$
(``independent'' lines in Fig. \ref{fig:sn3}) rather than the effect
of adding galaxy clustering information and the upper limits to $b$
and $r$ (``combined'').

We expect, on average, the signal-to-noise to decrease with redshift,
although there are more complicated trends on large scales (small $k$)
for $f_{{\rm g},mn}$ because galaxy samples at higher redshift probe
larger scales and their galaxy numbers increase first with redshift
followed by a decline beyond $z\sim0.9$. We have found that the most
significantly measured quantities are the galaxy clustering, then the
galaxy-matter cross-correlation, and then the matter clustering. In
reality, we should expect the significance of the first two to go
down, however, if an even smaller population of lenses is studied than
in the fiducial survey.

The signal-to-noise of the relative band power $f_{\delta,mn}$ drops
below \mbox{$\sim2\sigma$} on all scales beyond \mbox{$z\gtrsim0.4$}
so that we should not expect strong constraints beyond that
redshift. For a survey shallower than the fiducial survey, this limit
is bound to decrease yet further. The tightest constraints are
obtained for \mbox{$z\sim0.2$} and about \mbox{$k\sim2 h\,\rm
  Mpc^{-1}$} (\mbox{$\sim6\sigma$} detection), away from which the
detection fidelity drops relatively quickly below $3\sigma$ for
individual bands.  In conclusion, for \mbox{$z\lesssim0.8$} and
comoving \mbox{$0.5\lesssim k\,h^{-1}\,{\rm Mpc}\lesssim10$} the 3D
band power spectra $f_{{\delta\rm g},mn}$ and $f_{{\rm g},mn}$ can be
recovered most effectively, whereas a significant detection of
$f_{\delta,mn}$ is restricted to lower redshifts \mbox{$z\lesssim0.3$}
inside a range \mbox{$1\lesssim k h^{-1}{\rm Mpc}\lesssim10$}.  This
is in the heart of the regime discussed in
\citet{2011MNRAS.415.3649V}, hence an application of lensing
tomography promises tom improve our knowledge of the effect of baryons
on the matter power spectrum.

As we expect the noise covariances to be overly optimistic, a robust
study of the matter power spectrum would, however, probably require
coverage of a larger survey area than in our fiducial survey. We can
hope for a significant signal enhancement, if the survey area is
increased to $\sim10^3\,\rm deg^2$ or more, as expected from full-sky
surveys. The signal-to-noise should then roughly be boosted by
\mbox{$\sqrt{10}\approx3.3$} for all band power estimates. Moreover,
boosting the number density of sources, for example through
space-based surveys, also increases the signal-to-noise as shot-noise
on small scales is $\propto\overline{n}^{-1/2}$
\citep{2002A&A...396....1S}.

\subsection{Reconstructed galaxy biasing parameters}

\begin{figure}
  \begin{center}
    \epsfig{file=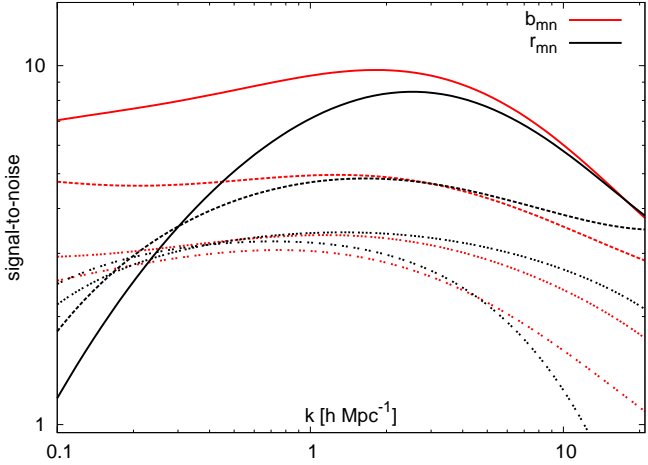,width=65mm,angle=-90}
   \end{center}
   \caption{\label{fig:sngb} Predicted signal-to-noise of the galaxy
     biasing parameters $(b_{mn},r_{mn})$ for different redshift bins
     $\bar{z}=0.15,0.4,0.6,0.85$ (from top to bottom line).}
\end{figure}

Combining the band power estimates, the Bayesian analysis provides
constraints on the galaxy biasing functions (see the two right-hand
panels in Fig. \ref{fig:bayeslogflat}). Here, the highest accuracy
(higher than $\sim3\sigma$) is achieved for $z\lesssim0.5$ and
\mbox{$1\lesssim k\,h^{-1}\,{\rm Mpc}\lesssim10$}
(Fig. \ref{fig:sngb}). The detection fidelity clearly depends on the
number density of lenses, for which here we assume the fiducial value
of $3\,\rm arcmin^{-2}$. An increase in the lens population size will
increase the number of galaxy pairs in the estimators for the angular
correlation functions and, therefore, reduce measurement noise. Since
$P_{\delta\rm g}(k,\chi)$ and $P_{\rm g}(k,\chi)$ are most easily
reconstructed, it may be sensible in a survey with poorly confined
$P_\delta(k,\chi)$ to utilise only those two to constrain
\mbox{$P_{\delta\rm g}/P_{\rm g}\propto r/b$} as a biasing function
that can be pinned down with the highest confidence.

The galaxy bias analysis here uses a flat prior with
\mbox{$|r(k,\chi)|\le2$} and \mbox{$b(k,\chi)\le4$}. The prior has the
effect that a highly-significant measurement of galaxy clustering
$f_{{\rm g},mn}$ will put a lower limit of $f_{{\rm g},mn}/b_{\rm
  high}^2\le f_{\delta,mn}$ on the matter clustering, where here
$b_{\rm high}=4$. For the same reason, $f_{\delta{\rm g},mn}$ imposes
a lower limit of $f_{{\rm g},mn}/(b_{\rm high}r_{\rm high})\le
f_{\delta,mn}$ for $r_{\rm high}=2$ to $f_{\delta,mn}$. Therefore,
combining galaxy clustering and shear tomography aids the recovery of
the matter power spectrum, as can be seen when comparing
``independent'' and ``combined'' lines in Fig. \ref{fig:sn3}. The
improvement is mostly present for low signal-to-noise bands in the
matter power spectrum. By varying the prior intervals of $b$ and $r$,
we found that the improvement is mainly due to the prior of $r$ rather
than $b$.

\subsection{Systematics}

The above predictions rely on the assumption that systematics in the
measured correlation functions are negligible. In reality, several
potential systematics -- unrelated to the data reduction and shape
measurements of galaxy images -- have been identified and several
points can be made:
\begin{itemize}

\item Correlations of intrinsic (unlensed) galaxy image shapes
  (``II'') add to the shear-shear correlation functions
  $\xi_\pm^{(ij)}$ \citep[e.g.,][]{Heymans06}.
\item The ``GI''-effect is more intricate \citep{2004PhRvD..70f3526H},
  as it is non-locally produced by correlations between the intrinsic
  source image shape and the shear effect of its surrounding matter on
  a more distant source.
\item Another unwanted effect that could arise is the influence of
  cosmic magnification
  \citep{1989ApJ...339L..53N,2001PhR...340..291B}. Magnification by
  matter density fluctuations in front of lenses changes the projected
  number density of lenses; to the lowest order to \mbox{$\kappa_{\rm
      g}\mapsto\kappa_{\rm g}+\lambda\kappa$}, where
  $\lambda:=2(\alpha-1)$ and \mbox{$\overline{N}_{\rm g}(>f)\propto
    f^{-\alpha}$} is the mean number density of lenses with fluxes
  greater than $f$, and $\kappa$ is the lensing convergence by
  intervening matter in front of a lens. This adds an additional
  contribution to $\overline{\gamma}^{(ij)}_{\rm t}(\Delta\theta)$ of
  the order of
  $\lambda\ave{\kappa(\theta)\kappa(\theta+\Delta\theta)}$ and a
  contribution to $\omega^{(ij)}(\Delta\theta)$ of the order of
  $\lambda^2\ave{\kappa(\theta)\kappa(\theta+\Delta\theta)}$. Since
  $\lambda$ is usually of the order of unity
  \citep{2010MNRAS.401.2093V}, both contaminations are of the order of
  the $\xi_+(\Delta\theta)$, and generated by matter in front of the
  lenses. In most cases, this signal is considerably weaker than the
  galaxy-galaxy lensing or lens clustering signal so that this effect
  is likely to be of only minor significance.
  
\end{itemize}
\begin{figure}
  \begin{center}
    \epsfig{file=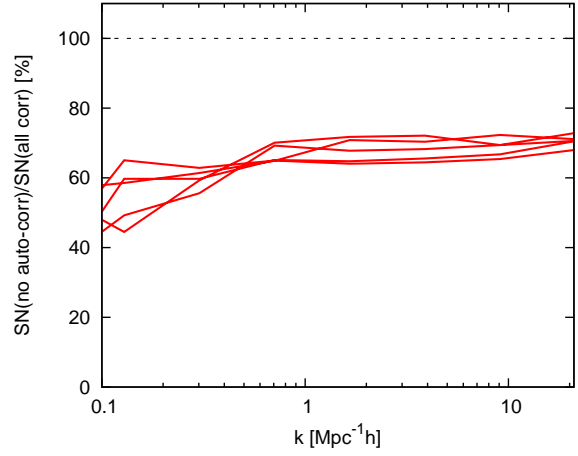,width=65mm,angle=-90}
   \end{center}
   \caption{\label{fig:sn2} Relative change in the signal-to-noise in
     $f_{\delta,mn}$ if only shear tomography correlation functions
     $\xi_\pm^{(ij)}$ with \mbox{$i\ne j$} (redshift slices with no
     overlap) are used for the estimation (minimum variance). The plot
     is based on a fiducial survey with \mbox{$\sigma_z=0$}. Different
     lines correspond to different $z$-bins. In total, roughly 30
     percent of the signal-to-noise is lost.}
\end{figure}
Although the II- and GI-contaminations are an obvious concern for this
kind of analysis, the focus of this paper is the performance of the
new estimators under ideal conditions. We therefore defer a
quantitative discussion of those effects to a future paper with the
comment that the contamination can be partly removed by considering
only slice pairs with little or no overlap in the tomography analysis
in $\xi_\pm^{(ij)}$. To achieve this, we redid the aforementioned
analysis for a fiducial survey with no redshift errors
(\mbox{$\sigma_z=0$}), i.e., and exactly no overlap for \mbox{$i\ne
  j$}, and compared the relative change in signal-to-noise of an
analysis that uses all tomography correlation functions to the
signal-to-noise in a analysis that utilises only the \mbox{$i\ne j$}
correlators. The impact of omitting that piece of information on the
significance of $\hat{\vec{f}}_\delta$ is presented by
Fig. \ref{fig:sn2}: The significance drops by approximately 30 percent
at all redshifts. The contaminations can possibly be more effectively
removed by means of nulling techniques \citep{2008A&A...488..829J}.
Since this is very likely to further degrade constraints on the 3D
power spectra, we can speculate with the predictions above that
significantly more survey area than $200\,\rm deg^2$ will then be
needed for a proper reconstruction of $P_\delta(k,\chi)$.

\subsection{B-mode matter band power spectrum}

Finally, the estimator for the 3D matter band power spectrum, based on
Eq. \Ref{eq:xipm}, presumes the absence of B-modes in the shear field,
i.e., that shear-shear correlations vanish after a rotation of the
source ellipticities $\epsilon$ by $\phi_{45}:=\pi/4$, hence after the
transformation $\epsilon\mapsto\e^{-2\i\phi_{45}}\epsilon$. The
presence of a B-mode component in $\xi_\pm^{(ij)}$ can potentially
distort $\vec{f}_\delta$. Whether such a contamination is present can
be verified in an analysis by estimating $\vec{f}^{\rm B}_\delta$ with
all source ellipticities rotated in the data by $\phi_{45}$ (B-mode
band power spectrum) and by estimating $\vec{f}^{\rm EB}_\delta$ with
only one source ellipticity in the two-point correlator
$\xi_\pm^{(ij)}$ rotated by $\phi_{45}$ (E/B-mode cross-power
spectrum). Hence, the inferred B-mode band powers $\vec{f}^{\rm
  B}_\delta$ and $\vec{f}^{\rm EB}_\delta$ that are inconsistent with
zero indicate the presence and magnitude of such a contamination. We
note that this only reveals possible B-mode issues in the data, but is
by no means a sufficient and necessary condition for systematics in
the data, since systematics could in principle only be present in the
E-mode.

To obtain an estimator for $P_\delta(k,\chi)$ that is purely rooted in
the E-mode components of the shear tomography correlations, it is also
conceivable to devise the COSEBI formalism, which was proposed by
\citet{2010A&A...520A.116S}. In this alternative framework, only the
E-mode components of the two-point shear correlation functions are
extracted. As the COSEBI values ${\rm E}_n$ are linear transformations
of $\xi_\pm(\theta)$ on a finite interval
\mbox{$\vartheta_0\le\theta\le\vartheta_1$}, they (a) can also be
easily measured from the data and (b) must also be linear
transformations of $\vec{f}_\delta$
\begin{equation}
  \vec{E}=\mat{V}\vec{f}_\delta\;,
\end{equation}
where the COSEBI statistics ${\rm E}^{(ij)}_n$ up to order $n=1\ldots
N_{\rm order}$ and pairs of redshift slices are assembled within the
new data vector $\vec{E}$. Therefore, one could use $\mat{V}$ instead
of $\mat{X}$ and $\vec{E}$ instead of $\vec{\xi}$ in the previous
estimator in Eq. \Ref{eq:xipmest} and the likelihood in
Eq. \Ref{eq:likelihood} to implement a matter band-power spectrum
reconstruction based on COSEBIs. Moreover, as shown by
\citet{2012arXiv1201.2669A} and \citet{2011MNRAS.418..536E}, to
estimate cosmological parameters with cosmic shear only a few COSEBIs
modes are required. If this also turns out to be the case for the
measurement of the band powers, this will dramatically reduce the size
of the noise covariances involved, possibly simplifying their
calculation or estimation from the data. It is unclear at this point,
however, up to which order $N_{\rm order}$ COSEBIs are required to
comprise essentially all information on $P_\delta(k,\chi)$.


\begin{acknowledgements}
  The author of this paper would like to thank Cristiano Porciani for
  useful discussions and, especially, Peter Schneider for discussions
  and his comments on the paper.  PS also acknowledges helpful
  comments by the reviewer of the paper, Ludovic van Waerbeke. This
  work was supported by the Deutsche Forschungsgemeinschaft in the
  framework of the Collaborative Research Center TR33 `The Dark
  Universe'. A big thanks also goes to Dennis Ritchie whose pioneering
  contributions to computer science made this work possible.
\end{acknowledgements}

\appendix

\section{Numerical evaluation of basis functions}
\label{sect:calculus}

To calculate the contribution of a power band to a two-point
correlator $\lambda_{mn}(\theta)$ on the sky, we note that all three
families of basis functions in Eqs. \Ref{eq:xipmbasis},
\Ref{eq:gammatbasis} and \Ref{eq:omegabasis} can be written in the
form
\begin{equation}
  \theta^2\lambda_{mn}(\theta)=
  \int_{\chi_n}^{\chi_{n+1}}\d\chi\,
  g(\chi)
  \int\limits_{k_mf_{\rm k}(\chi)\theta}^{k_{m+1}f_{\rm k}(\chi)\theta}\d s\,s\,J_n(s)\,
  P^{\rm fid}\left(\frac{s}{f_{\rm k}(\chi)\theta},\chi\right)\;.
\end{equation}
For a numerical evaluation of integrals of this type, we suggest
approximating the inner $s$-integral with the sum
\begin{eqnarray}
  \Lambda_{mn}(\theta,\chi)&:=&
  \int\limits_{k_mf_{\rm k}(\chi)\theta}^{k_{m+1}f_{\rm k}(\chi)\theta}\d s\,s\,J_n(s)\,
  P^{\rm fid}\left(\frac{s}{f_{\rm k}(\chi)\theta},\chi\right)\\
  \nonumber
  &\approx&
  \sum_{j=0}^{N_{\rm int}-1}P^{\rm fid}(\bar{k}_j,\chi)
  \int\limits_{k_jf_{\rm k}(\chi)\theta}^{k_{j+1}f_{\rm k}(\chi)\theta}\d s\,s\,
  J_n(s)\\
  \nonumber
  &=&
  \sum_{j=0}^{N_{\rm int}-1}P^{\rm fid}(\bar{k}_j,\chi)  
  \left[{\cal J}_n(\hat{k}_{j+1}f_{\rm k}(\chi)\theta)-
  {\cal J}_n(\hat{k}_jf_{\rm k}(\chi)\theta)\right]\;,
\end{eqnarray}
where 
\begin{equation}
  {\cal J}_n(x):=\left\{
    \begin{array}{ll}
      x J_1(x) & ,{\rm for}~n=0\\
      2-2 J_0(x)-x J_1(x) & ,{\rm for}~n=2\\
      4+\left(x-\frac{8}{x}\right) J_1(x)-8 J_2(x) & ,{\rm for}~n=4
    \end{array}\right.
\end{equation}
and $\bar{k}_j:=(\hat{k}_j+\hat{k}_{j+1})/2$, $\hat{k}_j:=k_m+j\Delta
k$ and $\Delta k:=(k_{m+1}-k_m)/N_{\rm int}$ for \mbox{$N_{\rm
    int}\sim100$} integration intervals. Where \mbox{$P^{\rm
    fid}(k,\chi)=1$} is a reference power spectrum, the foregoing sum
is exact for $N_{\rm int}=1$. The outer $\chi$-integral
\begin{equation}
  \theta^2\lambda_{mn}(\theta)=
  \int_{\chi_n}^{\chi_{n+1}}\d\chi\,g(\chi)
  \Lambda_{mn}(\theta,\chi)
\end{equation}
can be computed with Romberg's method \citep{1992nrca.book.....P}.

\section{Generation of mock catalogues}
\label{sect:montecarlo}

Here we outline an algorithm for making Monte Carlo realisations of
shear tomography data, including galaxy clustering and galaxy-galaxy
lensing. 

The convergence fields are simulated as Gaussian random fields,
whereas galaxy clustering obeys log-normal statistics
\citep{1991MNRAS.248....1C}. Log-normal statistics, or statistics with
a lower bound \mbox{$\kappa_{\rm g}\ge-1$} for the number density
contrast $\kappa_{\rm g}$, is essential for a catalogue of mock galaxy
positions. For the simulation, we use $2048\times2048\,\rm pixel^2$
grids, spanning an angular area of $4\times4\,\rm deg^2$, out of which
just $2\times 2\,\rm deg^2$ subfields are taken after performing our
realisation. The simulation comprises $N_z=5$ redshift slices. The
Monte Carlo recipe is similar to the approach outlined in
\citet{1988MNRAS.234..509C}, \cite{1989MNRAS.238..319C}, and
\citet{1991MNRAS.248....1C}, but generalised to the simultaneous
realisation of a set of random fields, both Gaussian and log-normal,
with defined cross-correlations. As a model of the 3D dark matter
power spectrum, the prescription of \citet{Smith03} is employed.

For the generation of correlated random fields, we follow the approach
of \citet{2004A&A...417..873S}, but now also include the clustering of
galaxies as additional degree of freedom.  A realisation is performed
on grids of angular size $A$, namely one lensing convergence grid
$\kappa^{(i)}(\vec{\theta})$ and one galaxy number density
$\kappa_{\rm g}^{(i)}(\vec{\theta})$ grid for each of the $i=1\ldots
N_z$ galaxy subsamples. The Fourier modes for identical $\vec{\ell}$
vectors of all $2N_z$ grids are compiled as one vector
\begin{equation}
  \tilde{\vec{v}}(\vec{\ell})=
  \left(
    \tilde{\kappa}^{(1)}(\vec{\ell}),\tilde{\kappa}^{(2)}(\vec{\ell}),\ldots,
    \tilde{\kappa}_{\rm g}^{(1)}(\vec{\ell}),\tilde{\kappa}_{\rm
      g}^{(2)}(\vec{\ell}),\ldots\right)^{\rm t}
\end{equation}
and are randomly generated independently from all other modes by
\citep[Sect. IV.B. in][]{2002PhRvD..66f3506H}
\begin{equation}
  \tilde{\vec{v}}(\vec{\ell})=
  \frac{1}{\sqrt{2A}}\mat{L}_S\left(\vec{g}+{\rm i}\vec{g}^\prime\right)\;,
\end{equation}
where $ g_i,g^\prime_i\curvearrowleft {\cal N}(0,1)$ are vectors of
normally distributed random numbers, with mean zero and unity
variance, having the same number of elements as
$\vec{\tilde{v}}(\vec{\ell})$. The matrix $\mat{L}_S$ denotes the Cholesky
decomposition, $\mat{S}=\mat{L}_S\mat{L}_S^{\rm t}$, of the Fourier
mode covariance matrix
\begin{equation}
  \label{eq:powermatrix}
  \mat{S}=
  \left(
    \begin{array}{ll}
      \mat{P}_{\kappa}(\ell) & \mat{P}_{\kappa\rm g}(\ell)\\
      \mat{P}_{\kappa\rm g}(\ell) & \mat{P}_{\rm g}(\ell)
    \end{array}
  \right)\;.
\end{equation}
This matrix $\mat{S}$ consists of three sub-matrices, describing (a)
the auto- and cross-correlations between the convergence fields by
\begin{equation}
  [\mat{P}_\kappa(\ell)]_{ij}=
  \frac{9H_0^4\Omega_{\rm m}^2}{4c^4}\int_0^{\chi_{\rm h}}
  \d\chi
    \frac{\overline{W}^{(i)}(\chi)\overline{W}^{(j)}(\chi)}
    {a(\chi)^2}
    P_\delta\left(\frac{\ell}{f_{\rm k}(\chi)},\chi\right)\;,
\end{equation}
(b) the lenses number density and convergence cross-correlations 
\begin{equation}
  [\mat{P}_{\kappa\rm g}(\ell)]_{ij}=
  \frac{3H_0^2\Omega_{\rm m}}{2c^2}\int_0^{\chi_{\rm h}}
  \d\chi
    \frac{\overline{W}^{(i)}(\chi)q_\chi^{(j)}(\chi)}
    {f_{\rm k}(\chi)a(\chi)}
  P_{\delta\rm g}\left(\frac{\ell}{f_{\rm k}(\chi)},\chi\right)\;,
\end{equation}
and (c) the auto- and cross-correlations between the lens number
density fields
\begin{equation}
  [\mat{P}_{\rm g}(\ell)]_{ij}=
  \hat{P}^{(ij)}_{\rm g}(\ell)\;,
\end{equation}
which needs further discussion below.  For the scope of this paper,
lenses are just random subsets of the source samples,
i.e. $p_\chi^{(i)}(\chi)=q_\chi^{(i)}(\chi)$. Moreover, galaxies are
assumed to be unbiased, i.e., $b(k,\chi)=r(k,\chi)=1$.

Case (c) in Eq. \Ref{eq:powermatrix} requires a special treatment to
ensure that a log-normal random field can be realised. We obtain
$\hat{P}^{(ij)}_{\rm g}(\ell)$ from
\begin{equation}
  P_{\rm g}^{(ij)}(\ell)= 
  \int_0^{\chi_{\rm h}}
  \d\chi
  \frac{q_\chi^{(i)}(\chi)q_\chi^{(j)}(\chi)}
  {f_{\rm k}^2(\chi)}
  P_{\rm g}\left(\frac{\ell}{f_{\rm k}(\chi)},\chi\right)
\end{equation}
in three steps:
\begin{eqnarray}
  {\rm step~1:} &
  \omega^{(ij)}(\theta)=
  \frac{1}{2\pi}\int_0^\infty\d\ell\,\ell\,
  J_0(\ell\theta)P_{\rm g}^{(ij)}(\ell)\;,\\
  {\rm step~2:} & 
  \hat{\omega}^{(ij)}(\theta)=
  \ln{(\omega^{(ij)}(\theta)+1)}\;,\\
  {\rm step~3:} &
  \hat{P}_{\rm g}^{(ij)}(\ell)=
  2\pi\int_0^\infty\d\theta\,\theta\,
  J_0(\ell\theta)\hat{\omega}^{(ij)}(\theta)\;.
\end{eqnarray}
This transformation rule is found because a log-normal random field
with lower limit $\kappa_{\rm ln}\ge-1$ can be generated from a random
Gaussian field $\kappa_{\rm g}$ via \citep{1991MNRAS.248....1C}
\begin{equation}
  \kappa_{\rm ln}(\vec{\theta})={\rm e}^{\kappa_{\rm g}(\vec{\theta})-\sigma^2_{\rm g}/2}-1\;,
\end{equation}
if the two-point correlation function of the underlying Gaussian
process is
\begin{equation}
  \omega_{\rm g}(\theta)=
  \Ave{\kappa_{\rm g}(\theta^\prime)
    \kappa_{\rm g}(\theta^\prime+\theta)}=
  \ln{\bigl(1+\omega_{\rm ln}(\theta)\bigr)}
\end{equation}
compared to the two-point correlations $\omega_{\rm ln}(\theta)$ in
the log-normal random field, where $\sigma_{\rm g}^2$ is the variance
in $\kappa_{\rm g}$. By the same argument, however, the cross-power
between galaxy number density and convergence grid in
$\mat{P}_{\kappa\rm g}(\ell)$ remains unchanged because
\begin{eqnarray}
  \lefteqn{\Ave{\kappa_{\rm ln}(\theta^\prime)
   \kappa_{\rm g}(\theta^\prime+\theta)}=}\\
&&\nonumber
 \Ave{\left({\rm e}^{\kappa_{\rm g}(\theta^\prime)-\sigma_{\rm g}^2/2}-1\right)
   \kappa_{\rm g}(\theta^\prime+\theta)}=
 \Ave{\kappa_{\rm g}(\theta^\prime)
   \kappa_{\rm g}(\theta^\prime+\theta)}\;, 
\end{eqnarray}
for
\begin{eqnarray}
  \Ave{f(\delta_1,\delta_2)}&=&
  \int\d\delta_1\d\delta_2P(\delta_1,\delta_2)f(\delta_1,\delta_2)\;,\\
  P(\delta_1,\delta_2)&=&\frac{1}{2\pi\sigma_1\sigma_2\sqrt{1-\psi(\theta)^2}}\,\times\\
  &&\nonumber  
    \exp{\left(-
        \frac{\delta_1^2\sigma_2^2+\delta_2^2\sigma_1^2-2\omega(\theta)\delta_1\delta_2}{2\sigma_1^2\sigma_2^2(1-\psi(\theta)^2)}\right)}\;,
\end{eqnarray}
where $\sigma_1^2=\Ave{\delta_1^2}$, $\sigma_2^2=\Ave{\delta_2^2}$ and
$\psi(\theta)=\omega(\theta)/(\sigma_1\sigma_2)$, and
$P(\delta_1,\delta_2)$ is a bivariate Gaussian p.d.f.

After all $\ell$-modes $\vec{\tilde{v}}(\vec{\ell})$ have been
realised, the grids are Fourier transformed to real space to yield the
convergence fields and the galaxy number density contrast fields on a
grid. The galaxy number density contrasts represent only the
underlying Gaussian fields, $\kappa_{\rm g}$, so far and hence
still need to be transformed into the log-normal density
contrasts. This is done by applying the additional mapping
$\kappa_{\rm ln}={\rm e}^{\kappa_{\rm g}-\sigma^2_{\rm g}/2}-1$
for every pixel. We acquire $\sigma_{\rm g}^2$ from the given
realisation of the Gaussian random field.

\begin{figure}
  \begin{center}
   \psfig{file=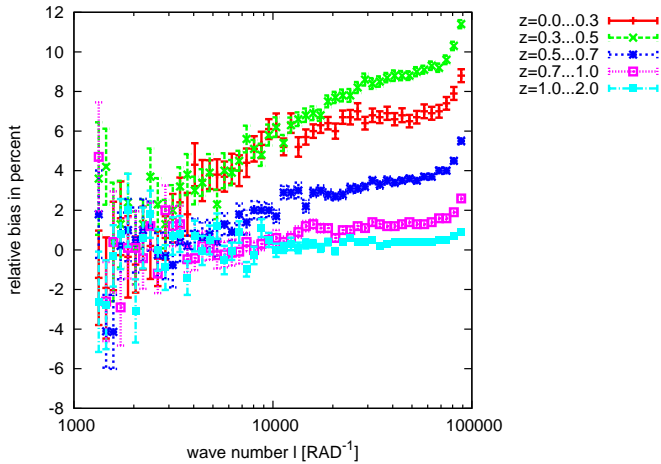,width=65mm,angle=-90}
 \end{center}
 \caption{\label{fig:lnbias} Relative bias $\Delta P_{\rm m}/P_{\rm m}^{\rm fid}$
   (percent) in the power spectrum of the log-normal random field realisation for
   different redshift bins as function of angular mode wave number $\ell$.}
\end{figure}

The mapping, however, is only an approximation because $\kappa_{\rm
  g}$ is actually a realisation of a \emph{smoothed} Gaussian field,
which is denoted here by an overline:
\begin{eqnarray}
  \overline{\kappa}_{\rm g}(\vec{\theta})&=&
  \int\d^2\theta^\prime\,
  \kappa_{\rm g}(\vec{\theta}^\prime)W(\vec{\theta}-\vec{\theta}^\prime)\;.
\end{eqnarray}
The function $W(\vec{\theta})$ represents the pixel smoothing
function, which is assumed to be normalised to unity by definition,
i.e. $\int\d^2\theta W(\vec{\theta})=1$. Applying the exponential
mapping to $\overline{\kappa}_{\rm g}$ thus assumes that a similar
relation between smoothed underlying Gaussian and smoothed log-normal
field $\overline{\kappa}_{\rm ln}$ approximately holds, namely
\begin{equation}
  \overline{\kappa}_{\rm ln}(\vec{\theta})\approx
  {\rm e}^{\overline{\kappa}_{\rm g}(\vec{\theta})-\sigma^2_{\rm g}/2}-1\;.
\end{equation}
The relation between the smoothed fields would be exact, if the
mapping were linear. The approximation has higher accuracy for less
fluctuating $\overline{\kappa}_{\rm g}$ and vice versa. For this
paper, we employ mock data in which galaxies perfectly trace the dark
matter distribution at all redshifts. As dark matter fluctuations
decline towards higher redshift, the above approximation provides a
gain in accuracy for higher redshift bins. With the outlined method,
we achieve an accuracy in the log-normal field power of better than
$10\%$ and on $\ell$-average several percent for the fiducial survey
employed (Fig. \ref{fig:lnbias}). We consider this to be sufficient
for our purposes, although a better treatment of this bias may be
desirable for future applications.

Finally, the grids $\kappa^{(i)}_{\rm g}$ and $\kappa^{(i)}$ serve as
a basis for the $i$th redshift slice of the mock survey: a candidate
galaxy position $\vec{\theta}$ is drawn randomly as the position on
the grid from a uniform PDF. The position is accepted, if
\mbox{$x\le1+\kappa^{(i)}_{\rm g}(\vec{\theta})$} where
\mbox{$x\in[0,1]$} is also a random number from a uniform PDF, in
accordance with \cite{1991MNRAS.248....1C} for grids large enough to
accommodate mostly no more than one galaxy per pixel. This algorithm
thus utilises a Poisson process to sample the galaxy number density
contrast $\kappa^{(i)}_{\rm g}$ with a discrete set of galaxy
positions. As a complex source ellipticity, we adopt
\mbox{$\gamma^{(i)}(\vec{\theta})+\epsilon_{\rm s}$}, where
$\gamma^{(i)}$ is the shear field corresponding to $\kappa^{(i)}$
\citep{1993ApJ...404..441K} and $\epsilon_{\rm s}$ the random
intrinsic shape of the galaxy drawn from a statistically independent
PDF. This process is repeated until the desired number density of
galaxies for the $i$th redshift bin is reached.

\bibliographystyle{aa}
\bibliography{invert}

\begin{thebibliography}{84}
\expandafter\ifx\csname natexlab\endcsname\relax\def\natexlab#1{#1}\fi

\bibitem[{{Asgari} {et~al.}(2012){Asgari}, {Schneider}, \&
  {Simon}}]{2012arXiv1201.2669A}
{Asgari}, M., {Schneider}, P., \& {Simon}, P. 2012, \texttt{arXiv:1201.2669}

\bibitem[{{Bacon} {et~al.}(2005){Bacon}, {Taylor}, {Brown}, \& {et
  al.}}]{2005MNRAS.363..723B}
{Bacon}, D.~J., {Taylor}, A.~N., {Brown}, M.~L., \& {et al.} 2005, \mnras, 363,
  723

\bibitem[{{Bartelmann} \& {Schneider}(2001)}]{2001PhR...340..291B}
{Bartelmann}, M. \& {Schneider}, P. 2001, \physrep, 340, 291

\bibitem[{{Baugh} \& {Efstathiou}(1994)}]{1994MNRAS.267..323B}
{Baugh}, C.~M. \& {Efstathiou}, G. 1994, \mnras, 267, 323

\bibitem[{{Blandford} {et~al.}(1991){Blandford}, {Saust}, {Brainerd}, \&
  {Villumsen}}]{1991MNRAS.251..600B}
{Blandford}, R.~D., {Saust}, A.~B., {Brainerd}, T.~G., \& {Villumsen}, J.~V.
  1991, \mnras, 251, 600

\bibitem[{{Bouchet} {et~al.}(1993){Bouchet}, {Strauss}, {Davis}, {Fisher},
  {Yahil}, \& {Huchra}}]{1993ApJ...417...36B}
{Bouchet}, F.~R., {Strauss}, M.~A., {Davis}, M., {et~al.} 1993, \apj, 417, 36

\bibitem[{{Castro} {et~al.}(2005){Castro}, {Heavens}, \&
  {Kitching}}]{2005PhRvD..72b3516C}
{Castro}, P.~G., {Heavens}, A.~F., \& {Kitching}, T.~D. 2005, \prd, 72, 023516

\bibitem[{{Clifton} {et~al.}(2011){Clifton}, {Ferreira}, {Padilla}, \&
  {Skordis}}]{2011arXiv1106.2476C}
{Clifton}, T., {Ferreira}, P.~G., {Padilla}, A., \& {Skordis}, C. 2011,
  \texttt{arXiv:1106.2476}

\bibitem[{{Clowe} {et~al.}(2006){Clowe}, {Brada{\v c}}, {Gonzalez}, \& {et
  al}}]{Clowe06_bullet}
{Clowe}, D., {Brada{\v c}}, M., {Gonzalez}, A.~H., \& {et al}. 2006, ApJ(Lett),
  648, L109

\bibitem[{{Coles}(1988)}]{1988MNRAS.234..509C}
{Coles}, P. 1988, \mnras, 234, 509

\bibitem[{{Coles}(1989)}]{1989MNRAS.238..319C}
{Coles}, P. 1989, \mnras, 238, 319

\bibitem[{{Coles} \& {Jones}(1991)}]{1991MNRAS.248....1C}
{Coles}, P. \& {Jones}, B. 1991, \mnras, 248, 1

\bibitem[{{Dodelson}(2003)}]{2003moco.book.....D}
{Dodelson}, S. 2003, {Modern cosmology}, ed. {Dodelson, S.}

\bibitem[{{Eifler}(2011)}]{2011MNRAS.418..536E}
{Eifler}, T. 2011, \mnras, 418, 536

\bibitem[{{Guzik} \& {Seljak}(2001)}]{2001MNRAS.321..439G}
{Guzik}, J. \& {Seljak}, U. 2001, \mnras, 321, 439

\bibitem[{{Hartlap} {et~al.}(2009){Hartlap}, {Schrabback}, {Simon}, \&
  {Schneider}}]{2009A&A...504..689H}
{Hartlap}, J., {Schrabback}, T., {Simon}, P., \& {Schneider}, P. 2009, \aap,
  504, 689

\bibitem[{{Hartlap} {et~al.}(2007){Hartlap}, {Simon}, \&
  {Schneider}}]{2007A&A...464..399H}
{Hartlap}, J., {Simon}, P., \& {Schneider}, P. 2007, \aap, 464, 399

\bibitem[{{Heitmann} {et~al.}(2010){Heitmann}, {White}, {Wagner}, {Habib}, \&
  {Higdon}}]{2010ApJ...715..104H}
{Heitmann}, K., {White}, M., {Wagner}, C., {Habib}, S., \& {Higdon}, D. 2010,
  \apj, 715, 104

\bibitem[{{Heymans} {et~al.}(2006){Heymans}, {White}, {Heavens}, {Vale}, \&
  {van Waerbeke}}]{Heymans06}
{Heymans}, C., {White}, M., {Heavens}, A., {Vale}, C., \& {van Waerbeke}, L.
  2006, MNRAS, 371, 750

\bibitem[{{Hilbert} {et~al.}(2011){Hilbert}, {Hartlap}, \&
  {Schneider}}]{2011arXiv1105.3980H}
{Hilbert}, S., {Hartlap}, J., \& {Schneider}, P. 2011, \aap, 536, A85

\bibitem[{{Hilbert} {et~al.}(2009){Hilbert}, {Hartlap}, {White}, \&
  {Schneider}}]{2009A&A...499...31H}
{Hilbert}, S., {Hartlap}, J., {White}, S.~D.~M., \& {Schneider}, P. 2009, \aap,
  499, 31

\bibitem[{{Hildebrandt} {et~al.}(2008){Hildebrandt}, {Wolf}, \&
  {Ben{\'{\i}}tez}}]{2008A&A...480..703H}
{Hildebrandt}, H., {Wolf}, C., \& {Ben{\'{\i}}tez}, N. 2008, \aap, 480, 703

\bibitem[{{Hirata} \& {Seljak}(2004)}]{2004PhRvD..70f3526H}
{Hirata}, C.~M. \& {Seljak}, U. 2004, Phys. Rev. D, 70, 063526

\bibitem[{{Hoekstra} {et~al.}(2002){Hoekstra}, {van Waerbeke}, {Gladders},
  {Mellier}, \& {Yee}}]{2002ApJ...577..604H}
{Hoekstra}, H., {van Waerbeke}, L., {Gladders}, M.~D., {Mellier}, Y., \& {Yee},
  H.~K.~C. 2002, \apj, 577, 604

\bibitem[{{Hu}(2002)}]{2002PhRvD..66h3515H}
{Hu}, W. 2002, \prd, 66, 083515

\bibitem[{{Hu} \& {Keeton}(2002)}]{2002PhRvD..66f3506H}
{Hu}, W. \& {Keeton}, C.~R. 2002, \prd, 66, 063506

\bibitem[{{Hu} \& {White}(2001)}]{2001ApJ...554...67H}
{Hu}, W. \& {White}, M. 2001, \apj, 554, 67

\bibitem[{{Jing} {et~al.}(2006){Jing}, {Zhang}, {Lin}, {Gao}, \&
  {Springel}}]{2006ApJ...640L.119J}
{Jing}, Y.~P., {Zhang}, P., {Lin}, W.~P., {Gao}, L., \& {Springel}, V. 2006,
  \apjl, 640, L119

\bibitem[{{Joachimi} \& {Schneider}(2008)}]{2008A&A...488..829J}
{Joachimi}, B. \& {Schneider}, P. 2008, \aap, 488, 829

\bibitem[{{Kaiser}(1992)}]{1992ApJ...388..272K}
{Kaiser}, N. 1992, \apj, 388, 272

\bibitem[{{Kaiser} \& {Squires}(1993)}]{1993ApJ...404..441K}
{Kaiser}, N. \& {Squires}, G. 1993, ApJ, 404, 441

\bibitem[{{Keitel} \& {Schneider}(2011)}]{2011arXiv1105.3672K}
{Keitel}, D. \& {Schneider}, P. 2011, \aap, 534, A76

\bibitem[{{Kiessling} {et~al.}(2011){Kiessling}, {Heavens}, {Taylor}, \&
  {Joachimi}}]{2011MNRAS.414.2235K}
{Kiessling}, A., {Heavens}, A.~F., {Taylor}, A.~N., \& {Joachimi}, B. 2011,
  \mnras, 414, 2235

\bibitem[{{Kitching} {et~al.}(2011){Kitching}, {Heavens}, \&
  {Miller}}]{2011MNRAS.413.2923K}
{Kitching}, T.~D., {Heavens}, A.~F., \& {Miller}, L. 2011, \mnras, 413, 2923

\bibitem[{{Komatsu} {et~al.}(2011){Komatsu}, {Smith}, \&
  {Dunkley}}]{2011ApJS..192...18K}
{Komatsu}, E., {Smith}, K.~M., \& {Dunkley}, e.~a. 2011, \apjs, 192, 18

\bibitem[{{Landy} \& {Szalay}(1993)}]{ls93}
{Landy}, S.~D. \& {Szalay}, A.~S. 1993, \apj, 412, 64

\bibitem[{{Laureijs} {et~al.}(2011){Laureijs}, {Amiaux}, {Arduini},
  {Augu{\`e}res}, {Brinchmann}, {Cole}, {Cropper}, {Dabin}, {Duvet}, {Ealet},
  \& et~al.}]{2011arXiv1110.3193L}
{Laureijs}, R., {Amiaux}, J., {Arduini}, S., {et~al.} 2011,
  \texttt{arXiv:1110.3193}

\bibitem[{{Leonard} {et~al.}(2012){Leonard}, {Dup{\'e}}, \&
  {Starck}}]{2011arXiv1111.6478L}
{Leonard}, A., {Dup{\'e}}, F.-X., \& {Starck}, J.-L. 2012, \aap, 539, A85

\bibitem[{{Li} {et~al.}(2011){Li}, {King}, {Zhao}, \&
  {Zhao}}]{2011MNRAS.415..881L}
{Li}, B., {King}, L.~J., {Zhao}, G.-B., \& {Zhao}, H. 2011, \mnras, 415, 881

\bibitem[{MacKay(2003)}]{2003Book...MACKAY}
MacKay, J.~D. 2003, Information Theory, Inference, and Learning Algorithms
  (Cambridge University Press, Cambridge, UK)

\bibitem[{{Mart\'{i}nez} \& {Saar}(2002)}]{2002sgd..book.....M}
{Mart\'{i}nez}, V.~J. \& {Saar}, E. 2002, {Statistics of the Galaxy
  Distribution}, ed. {Mart\'{i}nez, V.~J.~\& Saar, E.} (Chapman and Hall/CRC)

\bibitem[{{Massey} {et~al.}(2007{\natexlab{a}}){Massey}, {Rhodes}, \&
  {Ellis}}]{MasseyNat}
{Massey}, R., {Rhodes}, J., \& {Ellis}, R. 2007{\natexlab{a}}, Nat, 445, 286

\bibitem[{{Massey} {et~al.}(2007{\natexlab{b}}){Massey}, {Rhodes}, {Leauthaud},
  {Capak}, {Ellis}, {Koekemoer}, {R{\'e}fr{\'e}gier}, {Scoville}, {Taylor},
  {Albert}, {Berg{\'e}}, {Heymans}, {Johnston}, {Kneib}, {Mellier}, {Mobasher},
  {Semboloni}, {Shopbell}, {Tasca}, \& {Van Waerbeke}}]{2007ApJS..172..239M}
{Massey}, R., {Rhodes}, J., {Leauthaud}, A., {et~al.} 2007{\natexlab{b}},
  \apjs, 172, 239

\bibitem[{{Miralda-Escude}(1991)}]{1991ApJ...380....1M}
{Miralda-Escude}, J. 1991, \apj, 380, 1

\bibitem[{{Mo} {et~al.}(2010){Mo}, {van den Bosch}, \&
  {White}}]{2010gfe..book.....M}
{Mo}, H., {van den Bosch}, F.~C., \& {White}, S. 2010, {Galaxy Formation and
  Evolution}, ed. {Mo, H., van den Bosch, F.~C., \& White, S.}

\bibitem[{{Narayan}(1989)}]{1989ApJ...339L..53N}
{Narayan}, R. 1989, \apjl, 339, L53

\bibitem[{{Peacock}(1999)}]{1999coph.book.....P}
{Peacock}, J.~A. 1999, {Cosmological Physics}, ed. {Cambridge, UK: Cambridge
  University Press, ISBN 052141072X}

\bibitem[{{Peacock} \& {Dodds}(1996)}]{1996MNRAS.280L..19P}
{Peacock}, J.~A. \& {Dodds}, S.~J. 1996, \mnras, 280, L19

\bibitem[{{Peebles}(1980)}]{peebles80}
{Peebles}, P.~J.~E. 1980, {The large-scale structure of the universe}
  (Princeton University Press, USA)

\bibitem[{{Pen} {et~al.}(2003){Pen}, {Lu}, {van Waerbeke}, \&
  {Mellier}}]{2003MNRAS.346..994P}
{Pen}, U.-L., {Lu}, T., {van Waerbeke}, L., \& {Mellier}, Y. 2003, \mnras, 346,
  994

\bibitem[{{Pen} {et~al.}(2002){Pen}, {Van Waerbeke}, \&
  {Mellier}}]{2002ApJ...567...31P}
{Pen}, U.-L., {Van Waerbeke}, L., \& {Mellier}, Y. 2002, \apj, 567, 31

\bibitem[{{Press} {et~al.}(1992){Press}, {Teukolsky}, {Vetterling}, \&
  {Flannery}}]{1992nrca.book.....P}
{Press}, W.~H., {Teukolsky}, S.~A., {Vetterling}, W.~T., \& {Flannery}, B.~P.
  1992, {Numerical recipes in C. The art of scientific computing} (Cambridge:
  University Press, |c1992, 2nd ed.)

\bibitem[{{Schneider}(1998)}]{1998ApJ...498...43S}
{Schneider}, P. 1998, \apj, 498, 43

\bibitem[{{Schneider}(2006)}]{2006glsw.conf..269S}
{Schneider}, P. 2006, in Saas-Fee Advanced Course 33: Gravitational Lensing:
  Strong, Weak and Micro, ed. {G.~Meylan, P.~Jetzer, P.~North, P.~Schneider,
  C.~S.~Kochanek, \& J.~Wambsganss}, 269--451

\bibitem[{{Schneider} {et~al.}(2010){Schneider}, {Eifler}, \&
  {Krause}}]{2010A&A...520A.116S}
{Schneider}, P., {Eifler}, T., \& {Krause}, E. 2010, \aap, 520, 116

\bibitem[{{Schneider} \& {Hartlap}(2009)}]{2009A&A...504..705S}
{Schneider}, P. \& {Hartlap}, J. 2009, \aap, 504, 705

\bibitem[{{Schneider} {et~al.}(1998){Schneider}, {van Waerbeke}, {Jain}, \&
  {Kruse}}]{1998MNRAS.296..873S}
{Schneider}, P., {van Waerbeke}, L., {Jain}, B., \& {Kruse}, G. 1998, \mnras,
  296, 873

\bibitem[{{Schneider} {et~al.}(2002){Schneider}, {van Waerbeke}, {Kilbinger},
  \& {Mellier}}]{2002A&A...396....1S}
{Schneider}, P., {van Waerbeke}, L., {Kilbinger}, M., \& {Mellier}, Y. 2002,
  \aap, 396, 1

\bibitem[{{Schrabback} {et~al.}(2010){Schrabback}, {Hartlap}, \&
  {Joachimi}}]{schrabback2010}
{Schrabback}, T., {Hartlap}, J., \& {Joachimi}, B. 2010, \aap, 516, 63

\bibitem[{{Seljak} {et~al.}(2005){Seljak}, {Makarov}, {Mandelbaum}, {Hirata},
  {Padmanabhan}, {McDonald}, {Blanton}, {Tegmark}, {Bahcall}, \&
  {Brinkmann}}]{2005PhRvD..71d3511S}
{Seljak}, U., {Makarov}, A., {Mandelbaum}, R., {et~al.} 2005, \prd, 71, 043511

\bibitem[{{Semboloni} {et~al.}(2011){Semboloni}, {Hoekstra}, {Schaye}, {van
  Daalen}, \& {McCarthy}}]{2011arXiv1105.1075S}
{Semboloni}, E., {Hoekstra}, H., {Schaye}, J., {van Daalen}, M.~P., \&
  {McCarthy}, I.~G. 2011, \texttt{arXiv:1105.1075}

\bibitem[{{Semboloni} {et~al.}(2007){Semboloni}, {van Waerbeke}, {Heymans},
  {Hamana}, {Colombi}, {White}, \& {Mellier}}]{2007MNRAS.375L...6S}
{Semboloni}, E., {van Waerbeke}, L., {Heymans}, C., {et~al.} 2007, \mnras, 375,
  L6

\bibitem[{{Simon} {et~al.}(2007){Simon}, {Hetterscheidt}, {Schirmer}, {Erben},
  {Schneider}, {Wolf}, \& {Meisenheimer}}]{2007A&A...461..861S}
{Simon}, P., {Hetterscheidt}, M., {Schirmer}, M., {et~al.} 2007, \aap, 461, 861

\bibitem[{{Simon} {et~al.}(2012){Simon}, {Heymans}, {Schrabback}, {Taylor},
  {Gray}, {van Waerbeke}, {Wolf}, {Bacon}, {Barden}, {B{\"o}hm},
  {H{\"a}u{\ss}ler}, {Jahnke}, {Jogee}, {van Kampen}, {Meisenheimer}, \&
  {Peng}}]{2011arXiv1109.0932S}
{Simon}, P., {Heymans}, C., {Schrabback}, T., {et~al.} 2012, \mnras, 419, 998

\bibitem[{{Simon} {et~al.}(2004){Simon}, {King}, \&
  {Schneider}}]{2004A&A...417..873S}
{Simon}, P., {King}, L.~J., \& {Schneider}, P. 2004, \aap, 417, 873

\bibitem[{{Smith} {et~al.}(2003){Smith}, {Peacock}, {Jenkins}, {White},
  {Frenk}, {Pearce}, {Thomas}, {Efstathiou}, \& {Couchman}}]{Smith03}
{Smith}, R.~E., {Peacock}, J.~A., {Jenkins}, A., {et~al.} 2003, MNRAS, 341,
  1311

\bibitem[{{Somogyi} \& {Smith}(2010)}]{2010PhRvD..81b3524S}
{Somogyi}, G. \& {Smith}, R.~E. 2010, \prd, 81, 023524

\bibitem[{{Springel} {et~al.}(2005){Springel}, {White}, \&
  {Jenkins}}]{2005Natur.435..629S}
{Springel}, V., {White}, S.~D.~M., \& {Jenkins}, A. e.~a. 2005, \nat, 435, 629

\bibitem[{{Tegmark} {et~al.}(2004){Tegmark}, {Blanton}, {Strauss}, \& {etl
  a.}}]{2004ApJ...606..702T}
{Tegmark}, M., {Blanton}, M.~R., {Strauss}, M.~A., \& {etl a.} 2004, \apj, 606,
  702

\bibitem[{{Tegmark} \& {Peebles}(1998)}]{1998ApJ...500L..79T}
{Tegmark}, M. \& {Peebles}, P.~J.~E. 1998, \apjl, 500, 79

\bibitem[{{Tegmark} {et~al.}(1997){Tegmark}, {Taylor}, \&
  {Heavens}}]{1997ApJ...480...22T}
{Tegmark}, M., {Taylor}, A.~N., \& {Heavens}, A.~F. 1997, \apj, 480, 22

\bibitem[{{Tereno} {et~al.}(2009){Tereno}, {Schimd}, {Uzan}, {Kilbinger},
  {Vincent}, \& {Fu}}]{2009A&A...500..657T}
{Tereno}, I., {Schimd}, C., {Uzan}, J.-P., {et~al.} 2009, \aap, 500, 657

\bibitem[{{Uzan} \& {Bernardeau}(2001)}]{2001PhRvD..64h3004U}
{Uzan}, J.-P. \& {Bernardeau}, F. 2001, Phys. Rev. D, 64, 083004

\bibitem[{{van Daalen} {et~al.}(2011){van Daalen}, {Schaye}, {Booth}, \& {Dalla
  Vecchia}}]{2011MNRAS.415.3649V}
{van Daalen}, M.~P., {Schaye}, J., {Booth}, C.~M., \& {Dalla Vecchia}, C. 2011,
  \mnras, 415, 3649

\bibitem[{{van Waerbeke}(1998)}]{1998A&A...334....1V}
{van Waerbeke}, L. 1998, \aap, 334, 1

\bibitem[{{van Waerbeke}(2010)}]{2010MNRAS.401.2093V}
{van Waerbeke}, L. 2010, \mnras, 401, 2093

\bibitem[{{Van Waerbeke} \& {Mellier}(2003)}]{2003astro.ph..5089V}
{Van Waerbeke}, L. \& {Mellier}, Y. 2003, \texttt{arXiv:astro-ph/0305089}

\bibitem[{{VanderPlas} {et~al.}(2011){VanderPlas}, {Connolly}, {Jain}, \&
  {Jarvis}}]{2011ApJ...727..118V}
{VanderPlas}, J.~T., {Connolly}, A.~J., {Jain}, B., \& {Jarvis}, M. 2011, \apj,
  727, 118

\bibitem[{{Weinberg} {et~al.}(2004){Weinberg}, {Dav{\'e}}, {Katz}, \&
  {Hernquist}}]{2004ApJ...601....1W}
{Weinberg}, D.~H., {Dav{\'e}}, R., {Katz}, N., \& {Hernquist}, L. 2004, \apj,
  601, 1

\bibitem[{{White} \& {Hu}(2000)}]{2000ApJ...537....1W}
{White}, M. \& {Hu}, W. 2000, \apj, 537, 1

\bibitem[{{Yoshikawa} {et~al.}(2001){Yoshikawa}, {Taruya}, {Jing}, \&
  {Suto}}]{2001ApJ...558..520Y}
{Yoshikawa}, K., {Taruya}, A., {Jing}, Y.~P., \& {Suto}, Y. 2001, \apj, 558,
  520

\bibitem[{{Zaroubi} {et~al.}(1995){Zaroubi}, {Hoffman}, {Fisher}, \&
  {Lahav}}]{1995ApJ...449..446Z}
{Zaroubi}, S., {Hoffman}, Y., {Fisher}, K.~B., \& {Lahav}, O. 1995, \apj, 449,
  446

\bibitem[{{Zehavi} {et~al.}(2011){Zehavi}, {Zheng}, {Weinberg}, \& {et
  al.}}]{2011ApJ...736...59Z}
{Zehavi}, I., {Zheng}, Z., {Weinberg}, D.~H., \& {et al.} 2011, \apj, 736, 59

\bibitem[{{Zhan} \& {Knox}(2004)}]{2004ApJ...616L..75Z}
{Zhan}, H. \& {Knox}, L. 2004, \apjl, 616, L75

\end{thebibliography}

\end{document}